\documentclass[twocolumn]{aastex61}

\usepackage{multirow}

\begin{document}
\title{ARES\footnote{ARES: Ariel Retrieval of Exoplanets School} III: Unveiling the Two Faces of KELT-7\,b with HST WFC3}

\received{1 April 2020}
\revised{29 May 2020}
\accepted{23 June 2020}

\correspondingauthor{William Pluriel}
\email{william.pluriel@u-bordeaux.fr}

\author{William Pluriel}  
\affil{Laboratoire d'astrophysique de Bordeaux, Univ. Bordeaux, CNRS, B18N, all\'{e}e Geoffroy Saint-Hilaire, 33615 Pessac, France}

\author{Niall Whiteford}
 \affil{Institute for Astronomy, University of Edinburgh, Blackford Hill, Edinburgh, EH9 3HJ, UK}
    \affil{Centre for Exoplanet Science, University of Edinburgh, Edinburgh, EH9 3FD, UK}

\author{Billy Edwards}
\affil{Department of Physics and Astronomy, University College London, London, WC1E 6BT, UK}

\author{Quentin Changeat}
\affil{Department of Physics and Astronomy, University College London, London, WC1E 6BT, UK}

\author{Kai Hou Yip}
\affil{Department of Physics and Astronomy, University College London, London, WC1E 6BT, UK}

\author{Robin Baeyens}
\affil{Instituut voor Sterrenkunde, KU~Leuven, Celestijnenlaan 200D bus 2401, 3001 Leuven, Belgium}

\author{Ahmed Al-Refaie} 
\affil{Department of Physics and Astronomy, University College London, London, WC1E 6BT, UK}

\author{Michelle Fabienne Bieger}
\affil{College of Engineering, Mathematics and Physical Sciences, Physics Building, University of Exeter, North Park Road, Exeter, UK}

\author{Dorian Blain}
 	\affil{LESIA, Observatoire de Paris, Université PSL, CNRS, Sorbonne Universit\'e, Universit\'e de Paris, 5 place Jules Janssen, 92195 Meudon, France}

\author{Am\'elie Gressier}
\affil{Sorbonne Universit\'es, UPMC Universit\'e Paris 6 et CNRS, 
		UMR 7095, Institut d'Astrophysique de Paris, 98 bis bd Arago,
		75014 Paris, France}
 	\affil{LESIA, Observatoire de Paris, Université PSL, CNRS, Sorbonne Universit\'e, Universit\'e de Paris, 5 place Jules Janssen, 92195 Meudon, France}

\author{Gloria Guilluy}
	\affil{Dipartimento di Fisica, Universit\'{a} degli Studi di Torino, via Pietro Giuria 1, I-10125 Torino, Italy}
	\affil{INAF Osservatorio Astrofisico di Torino, Via Osservatorio 20, I-10025 Pino Torinese, Italy}

\author{Adam Yassin Jaziri}  
\affil{Laboratoire d'astrophysique de Bordeaux, Univ. Bordeaux, CNRS, B18N, all\'{e}e Geoffroy Saint-Hilaire, 33615 Pessac, France}
	
\author{Flavien Kiefer}
\affil{Sorbonne Universit\'es, UPMC Universit\'e Paris 6 et CNRS, 
		UMR 7095, Institut d'Astrophysique de Paris, 98 bis bd Arago,
		75014 Paris, France}

\author{Darius Modirrousta-Galian} 
	\affil{INAF – Osservatorio Astronomico di Palermo, Piazza del Parlamento 1, I-90134 Palermo, Italy}
	\affil{University of Palermo, Department of Physics and Chemistry, Via Archirafi 36, Palermo, Italy}

\author{Mario Morvan}
	\affil{Department of Physics and Astronomy, University College London, London, WC1E 6BT, UK}

\author{Lorenzo V. Mugnai} 
	\affil{La Sapienza Universit\'a di Roma, Department of Physics, Piazzale Aldo Moro 2, 00185 Roma, Italy}

\author{Mathilde Poveda}
	\affil{Laboratoire Interuniversitaire des Syst\`{e}mes Atmosph\'{e}riques (LISA), UMR CNRS 7583, Universit\'{e} Paris-Est-Cr\'eteil, Universit\'e de Paris, Institut Pierre Simon Laplace, Cr\'{e}teil, France}
	\affil{Maison de la Simulation, CEA, CNRS, Univ. Paris-Sud, UVSQ, Universit\'{e} Paris-Saclay, F-91191 Gif-sur-Yvette, France} 

 \author{Nour Skaf} 
 	\affil{LESIA, Observatoire de Paris, Université PSL, CNRS, Sorbonne Universit\'e, Universit\'e de Paris, 5 place Jules Janssen, 92195 Meudon, France}
	 \affil{Department of Physics and Astronomy, University College London, London, WC1E 6BT, UK}

 \author{Tiziano Zingales} 
 	\affil{Laboratoire d'astrophysique de Bordeaux, Univ. Bordeaux, CNRS, B18N, all\'{e}e Geoffroy Saint-Hilaire, 33615 Pessac, France}

\author{Sam Wright}
\affil{Department of Physics and Astronomy, University College London, London, WC1E 6BT, UK}

\author{Benjamin Charnay}
 	\affil{LESIA, Observatoire de Paris, Université PSL, CNRS, Sorbonne Universit\'e, Universit\'e de Paris, 5 place Jules Janssen, 92195 Meudon, France}

\author{Pierre Drossart}  
 	\affil{LESIA, Observatoire de Paris, Université PSL, CNRS, Sorbonne Universit\'e, Universit\'e de Paris, 5 place Jules Janssen, 92195 Meudon, France}

\author{J\'{e}r\'{e}my Leconte}  
\affil{Laboratoire d'astrophysique de Bordeaux, Univ. Bordeaux, CNRS, B18N, all\'{e}e Geoffroy Saint-Hilaire, 33615 Pessac, France}

\author{Angelos Tsiaras}
\affil{Department of Physics and Astronomy, University College London, London, WC1E 6BT, UK}

\author{Olivia Venot}  
\affil{Laboratoire Interuniversitaire des Syst\`{e}mes Atmosph\'{e}riques (LISA), UMR CNRS 7583, Universit\'{e} Paris-Est-Cr\'eteil, Universit\'e de Paris, Institut Pierre Simon Laplace, Cr\'{e}teil, France}

\author{Ingo Waldmann}
\affil{Department of Physics and Astronomy, University College London, London, WC1E 6BT, UK}

\author{Jean-Philippe Beaulieu}
\affil{School of Physical Sciences, University of Tasmania,
Private Bag 37 Hobart, Tasmania 7001 Australia}
\affil{Sorbonne Universit\'es, UPMC Universit\'e Paris 6 et CNRS, UMR 7095, Institut d'Astrophysique de Paris, 98 bis bd Arago, 75014 Paris, France}

	
\begin{abstract}

We present the analysis of the hot-Jupiter KELT-7\,b using transmission and emission spectroscopy from the Hubble Space Telescope (HST), both taken with the Wide Field Camera 3 (WFC3). Our study uncovers a rich transmission spectrum which is consistent with a cloud-free atmosphere and suggests the presence of H$_2$O and H-. In contrast, the extracted emission spectrum does not contain strong absorption features and, although it is not consistent with a simple blackbody, it can be explained by a varying temperature-pressure profile, collision induced absorption (CIA) and H-. KELT-7\,b had also been studied with other space-based instruments and we explore the effects of introducing these additional datasets. Further observations with Hubble, or the next generation of space-based telescopes, are needed to allow for the optical opacity source in transmission to be confirmed and for molecular features to be disentangled in emission.


\end{abstract}


\section{Introduction} \label{sec:intro}

With thousands of planets detected during the previous two decades, the study of atmospheres is at the forefront of exoplanet research and spectroscopic observations now probe these worlds in search of molecular features. Such studies are crucial in the pursuit of understanding the diverse nature of exoplanet chemical compositions, atmospheric processes, internal structures and the conditions required for planetary formation.

In recent years, there has been a surge in transit spectroscopy observations using both space-borne and ground-based facilities, resulting in significant advancements in our understanding of exoplanet atmospheres. This technique has been used for the detection of multiple molecular absorption features, including water (H$_2$O, \citet{tinetti_water, k2_18b}), methane (CH$_4$, \citet{swain_methane}) and ammonia \citep{macdonald_hd209}, becoming a cornerstone technique in the pursuit of exoplanet characterisation. In particular, the Hubble Space Telescope (HST) has been widely used allowing us to characterise the atmospheres of several hot Jupiters \citep[e.g.][]{Wakeford2013,Deming2013,Haynes_2015,Kreidberg_2018,Evans_2019} and has begun to observe enough planets for population studies to be undertaken \citep[e.g.][]{sing,angelos30}.

For hotter planets, HST transmission spectroscopy has provided evidence for absorption at shorter wavelengths. These are generally attributed to the presence of optical absorbers such as Titanium Oxide (TiO), Vanadium Oxide (VO) or Iron Hydride (FeH). These planets include WASP-121b \citep{Evans_2019} and  WASP-127\,b \citep{ares2}. Additionally, high-resolution ground-based observations have detected the presence of a variety of heavy metals in the atmosphere of KELT-9\,b \citep{Hoeijmaker_2018} while lower resolution data were used to claim the presence of Aluminium Oxide (AlO) in WASP-33\,b \citep{essen_wasp33}. 
Thermal emission spectra can also be used to characterise the day side of exoplanets. In particular, hot-Jupiters which emit significant and detectable infrared radiation are the ideal candidates for this approach. As thermal emission is sensitive to the temperature pressure profile \citep{Madhusudhan_2014_exoplanetaryatm}, it is possible to probe this structure and demonstrate how it can be driven by the presence of TiO in the atmosphere such as in the cases of WASP-33\,b \citep{Haynes_2015} and WASP-76\,b \citep{ares1}.

The presence of these optical absorbers are an important component in current exoplanet atmospheric models, and are predicted to have a significant impact on the overall physics and chemistry of highly irradiated close-in giant planet atmospheres. Work by \cite{Fortney_2008} suggests the day side atmospheres of such planets can be divided into two separate classes. The first class represents atmospheres which are modelled to have significant opacity due to the presence of TiO and VO gases and thus results in, for example, atmospheric temperature inversions (with hot stratospheres). In the second case, atmospheres which do not possess opacity due TiO and VO, are modelled to redistribute absorbed energy more readily resulting in cooler day-sides and warmer night-sides.

WASP-103\,b, a similar planet of KELT-7\,b, has been studied in both emission and transmission by HST, suggesting a thermal inversion on the dayside but with a featureless transmission spectrum \citep{Kreidberg_2018}. On the other hand, HST data for WASP-76\,b, another ultra-hot Jupiter, also showed a dayside thermal inversion due to the presence of TiO but a larger water feature and no evidence for optical absorbers in the transmission spectrum \citep{ares1}.

KELT-7\,b is a transiting hot-Jupiter, with mass of $1.28^{+0.18}_{-0.18} $M$_J$, radius $1.533^{+0.046}_{-0.047} $R$_J$, orbital period 2.7347749 $\pm$ 0.0000039 days and equilibrium temperature $\sim$2048 K \citep{Bieryla_2015} (see Table \ref{tab:planet_para}). KELT-7\,b, with relatively low surface gravity, high equilibrium temperature, and a bright host star, is an excellent candidate for thorough atmospheric characterisation. It has previously been studied in both transmission and emission using the Spitzer Space Telescope's InfraRed Array Camera (IRAC) \citep{garhart_spitzer} and a ground-based eclipse was measured by \citet{martioli_k7}. Here we present an analysis of this exoplanet using transmission and emission spectroscopy from the Hubble Space Telescope's WFC3. These HST observations allow for two complementary insights into the nature of this planet. We also explore the effects of combining the HST dataset with those from Spitzer IRAC and TESS. 

\section{Data analysis} \label{data_analysis}

\subsection{Observational Data - HST}

We obtained the raw spectroscopic observation data from the Mikulski Archive for Space Telescopes\footnote{\url{https://archive.stsci.edu/}}. Both observations of KELT-7\,b were undertaken in 2017 as part of Hubble proposal 14767 led by David Sing. These images are the result of a two visits of the target, each containing five HST orbits, using the infrared detector, the Wide Field Camera 3 (WFC3) G141 grism, and a scan rate of 0.9 s$^{-1}$. Each image consists of five non-destructive reads with an aperture size of 266 $\times$ 266 pixels in the SPARS10 mode, resulting in a total exposure time of 22.317s, a maximum signal level of 33,000 e$^{-}$ per pixel, and a total scan length of about 25.605 arcsec.

\subsection{Reduction and Extraction - Iraclis}

The data reduction and calibration was performed using the open-source software Iraclis \citep{Iraclis}. Our Iraclis analysis starts with the raw spatially scanned spectroscopic images. The data reduction and correction steps are performed in the following order: zero-read subtraction; reference pixel correction; non-linearity correction; dark current subtraction; gain conversion; sky background subtraction; calibration; flat-field correction; bad pixels and cosmic ray correction. Detailed description of the data reduction process could be found in Section 2 of \cite{Iraclis}.

\begin{figure}
    \centering
    \includegraphics[width=\columnwidth]{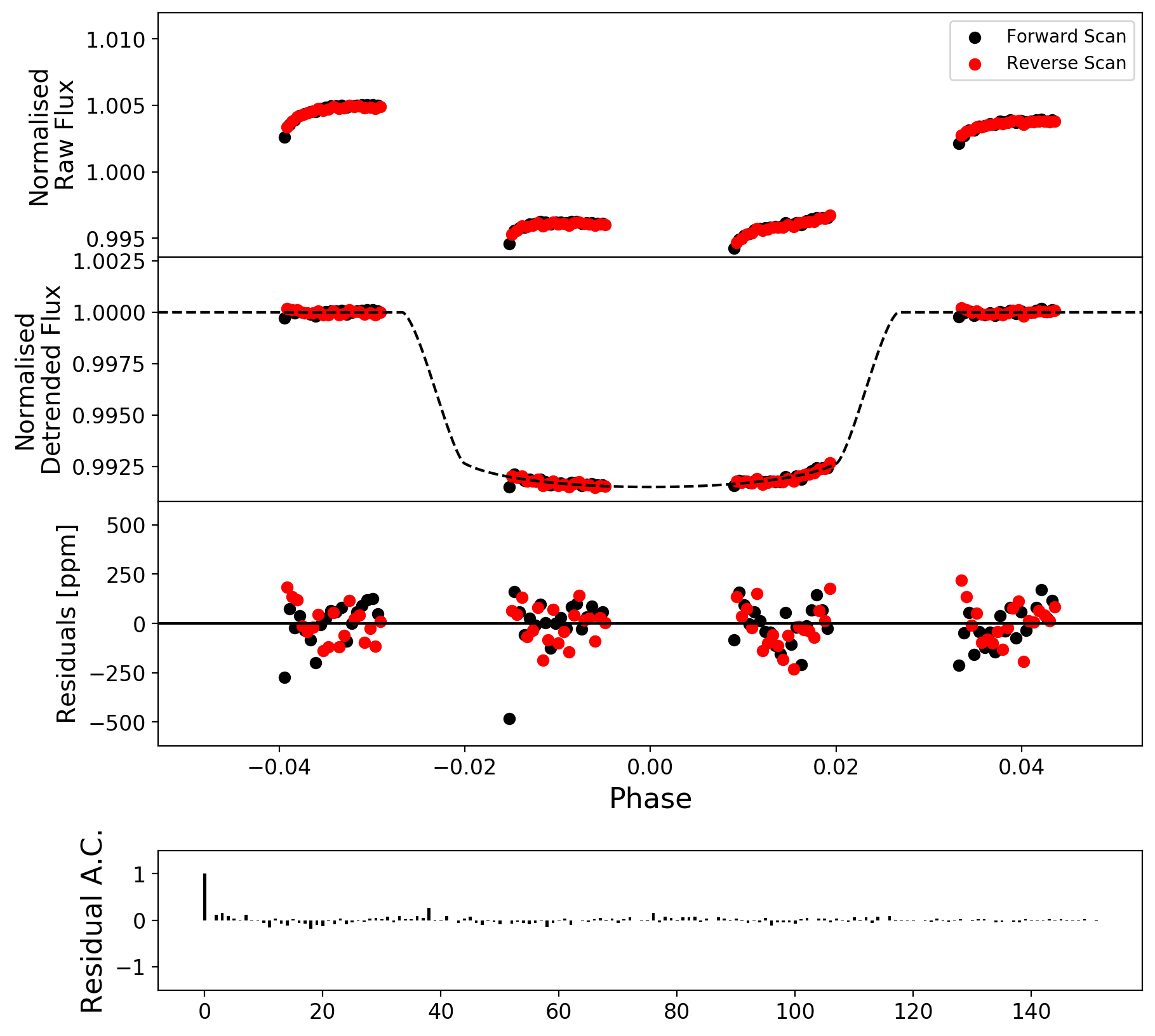}
    \includegraphics[width=\columnwidth]{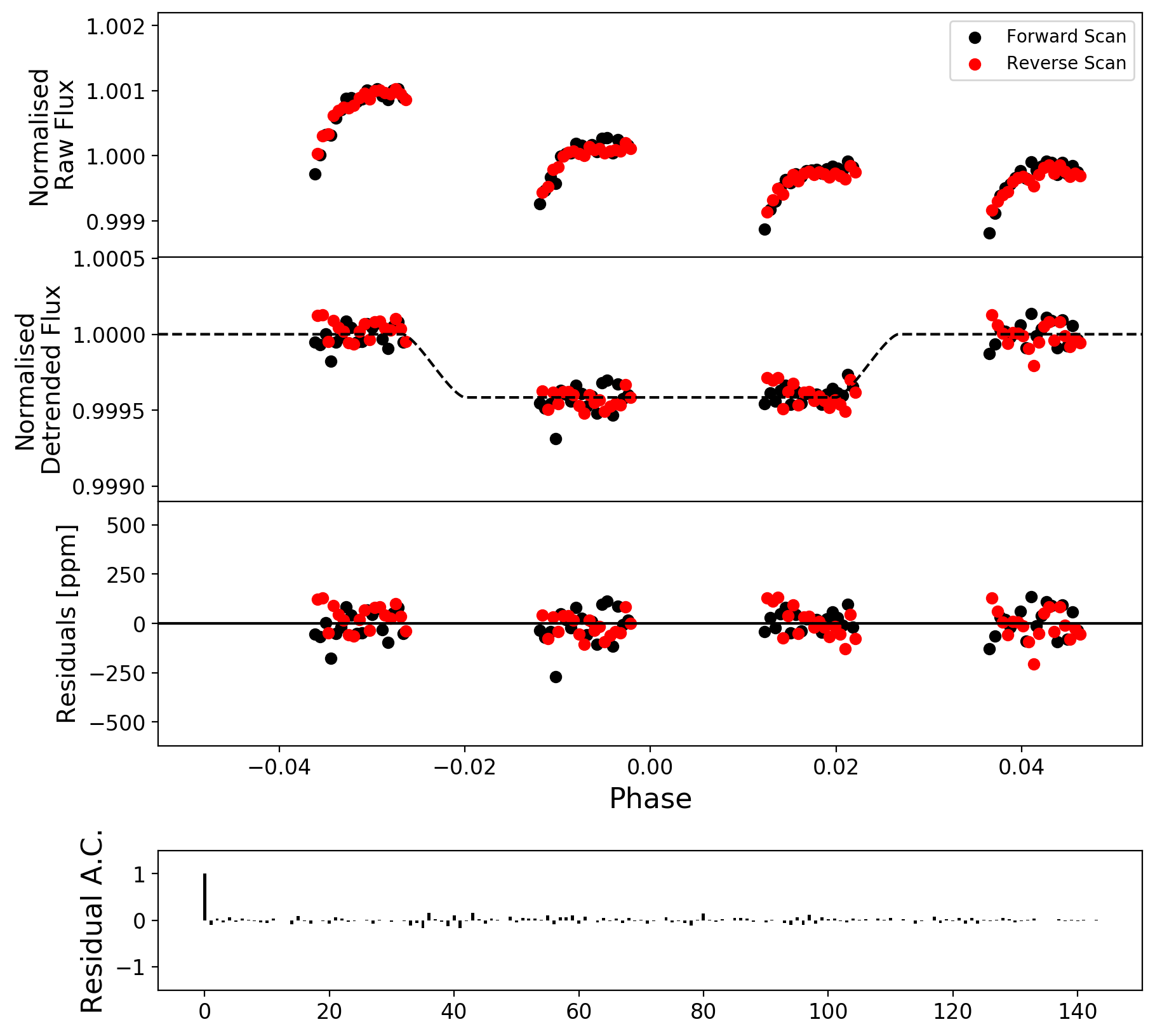}
    \caption{White light-curve for transmission (top) and emission (bottom) observations of KELT-7\,b. For each panel; Top: raw light-curve, after normalisation. Second: light-curve, divided by the best fit model for the systematics. Third: residuals. Bottom: auto-correlation function of the residuals.}
    \label{fig:white}
\end{figure}

\begin{figure*}
    \centering
    \includegraphics[width=0.475\textwidth]{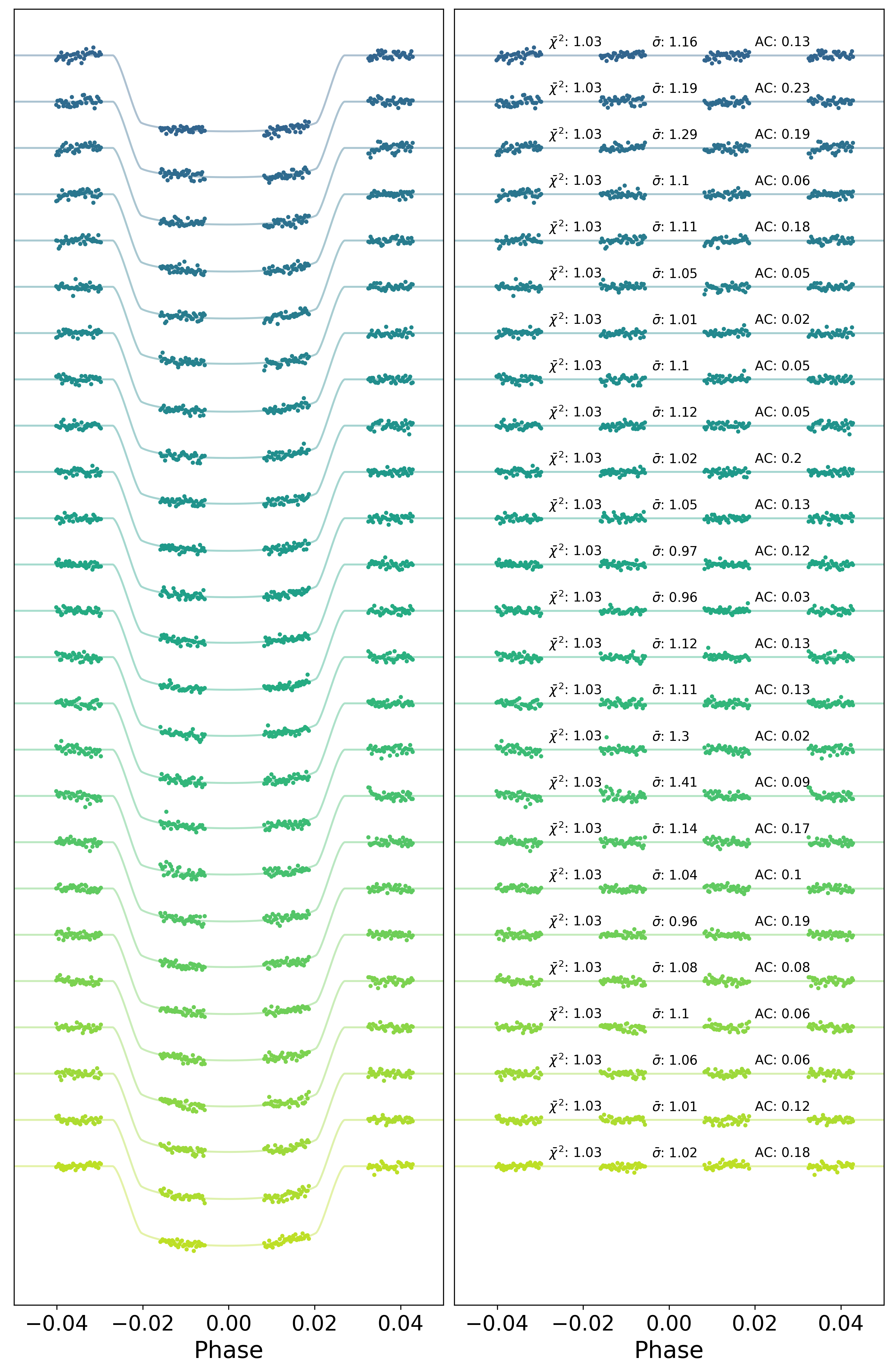}
    \includegraphics[width=0.475\textwidth]{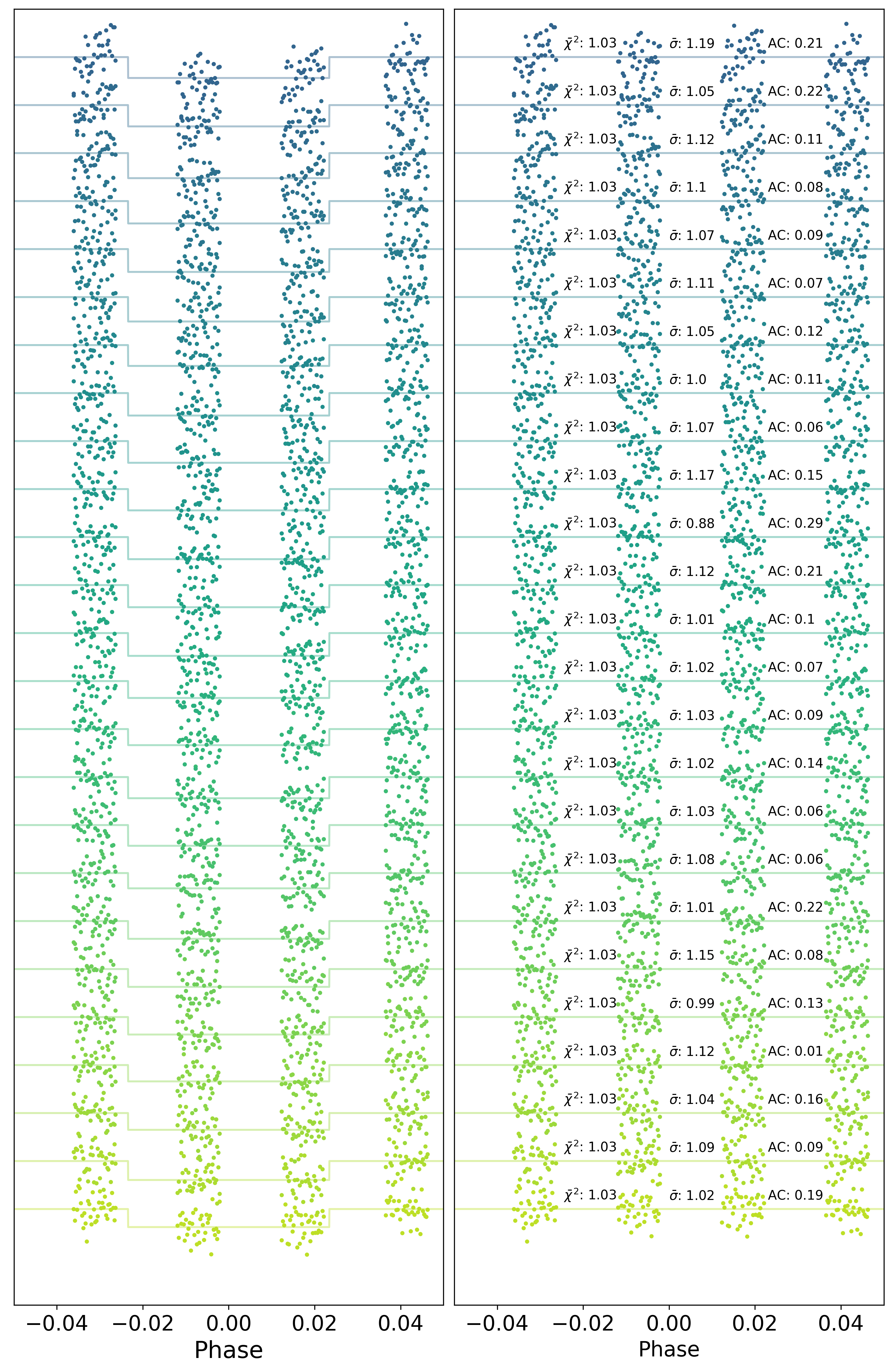}
    \caption{Spectral light-curve fits from Iraclis of transmission (left) and emission (bright) spectra - for clarity, an offset has been applied. In each plot, the left segment shows the spectral light-curves while the residuals are shown in the right section. $\overline{\chi}$ is the reduced chi squared, $\overline{\sigma}$ is the ratio between the standard deviation of the residuals and the photon noise, and AC is the auto-correlation of the fitting residuals.}
    \label{fig:all}
\end{figure*}{}

Following the reduction process, the flux was extracted from the spatially scanned spectroscopic images to create the final transit light-curves per wavelength band. We considered one broadband (white) light-curve covering the whole wavelength range in which the G141 grism is sensitive (1.088-1.68$\mu$m) and spectral light-curves with a resolving power of 70 at 1.4$\mu$m. The bands of the spectral light-curves are selected such that the SNR is approximately uniform across the planetary spectrum. We extracted our final light-curves from the differential, non-destructive reads. Prior to light curve fitting, we choose to discard the first HST orbit of each visit, as these exhibit much stronger hooks than subsequent orbits \citep{Deming2013, Zhou_2017}.

Our white and spectral light-curves were fit using the literature values from Table ~\ref{tab:retrieval}, with only two free parameters: the planet-to-star radius ratio and the mid-transit time. This is motivated by the Earth obscuration gaps, which often means the ingress and egress of the transit are missed, limiting our ability to refine the semi-major axis to star radius ratio, the inclination and the eccentricity. The limb-darkening coefficients were selected from the quadratic formula by \citet{Claret}, and using the stellar parameters in Table \ref{tab:planet_para} and the ATLAS stellar models \citep{Kurucz_1970_ATLAS,Espinoza2015,morello_exotethys}. Figure \ref{fig:white} shows the raw white light-curve, the detrended white light-curve, and the fitting residuals for both observations while Figure \ref{fig:all} shows the fits of spectral light-curves for each wavelength.

\begin{table}
\centering
\resizebox{\columnwidth}{!}{
\begin{tabular}{ll} 
	\hline\hline
	\multicolumn{2}{c}{Stellar parameters} \\
	\hline
	Spectral type & F \\
	Effective Temperature [K] & $\mathrm{6789}^{+50}_{-49}$ \\
	$[$Fe/H$]$ & $\mathrm{0.139}^{+.075}_{-0.081} $ \\
	Surface gravity [cgs] & $\mathrm{4.149}^{+0.019}_{-0.019}$ \\
	Radius [R$_{\odot}$] & $\mathrm{1.732}^{+0.043}_{-0.045} $ \\ 
	\hline
	\multicolumn{2}{c}{Planetary parameters} \\
	\hline
	Period [day] & $ \mathrm{2.7347749}^{+0.0000039}_{-0.0000039} $ \\
	Inclination [deg] & $ \mathrm{83.76}^{+0.38}_{-0.37} $ \\
	Mass [M$_{\mathrm{J}}$] & $ \mathrm{1.28}^{+0.18}_{-0.18} $ \\
	Radius [R$_{\mathrm{J}}$] & $ \mathrm{1.533}^{+0.046}_{-0.047} $ \\
	Equilibrium Temperature [K] & $ \mathrm{2048}^{+27}_{-27} $ \\
	T$_{0}$[BJD$_{\mathrm{TBD}}$] & 2456355.229809$^{+0.000198}_{-0.000198}$ \\
	a/R$_{*}$ & 5.49$^{+0.12}_{-0.11}$ \\
	\hline
	Reference & \citet{Bieryla_2015} \\
	\hline\hline
\end{tabular}
\label{tab:planet_para}
}
\caption{Target parameters used in this study.}
\end{table}

\subsection{Atmospheric characterisation - TauREx3} \label{sec:atmochar}

The reduced spectra obtained using Iraclis were thereafter fitted using TauREx3 \citep{al-refaie_taurex3}, a publicly available\footnote{\url{https://github.com/ucl-exoplanets/TauREx3_public}} Bayesian retrieval framework. For the star parameters and the planet mass, we used the values from \cite{Bieryla_2015} listed in Table \ref{tab:planet_para}. In our runs we assumed that KELT-7\,b possesses a primary atmosphere with a ratio H$_2$/He = 0.17. For the opacity sources, we used the line lists from the ExoMol project \citep{ExoMol}, along with those from HITRAN and HITEMP \citep{HITRAN,HITEMP}. In this publication, we considered six trace gases: H$_2$O \citep{polyansky_h2o}, CO\citep{li_co_2015}, TiO \citep{McKemmish_TiO_new}, VO \citep{mckemmish_vo}, FeH \citep{dulick_FeH,wende_FeH} and H- \citep{john_1988_h-,ares1}. For H-, the absorption depends only on the mixing ratios of neutral hydrogen atoms and free elections. As this is a degenerate problem, we fixed the neutral hydrogen volume mixing ratio and imposed a profile inspired from \cite{Parmentier_2018} using the two-layer model from \cite{le2layer}. The only remaining free parameter is log(e-). Clouds are modelled assuming a fully opaque grey opacity model.

In this study we use the plane-parallel approximation to model the atmospheres, with pressures ranging from $10^{-2}$ to $10^6$ Pa, uniformly sampled in log-space with 100 atmospheric layers. We included the Rayleigh scattering and the collision induced absorption (CIA) of H$_2$--H$_2$ and H$_2$--He \citep{abel_h2-h2, fletcher_h2-h2, abel_h2-he}. Constant molecular abundance profiles were used, and allowed to vary freely between $10^{-12}$ and $10^{-1}$ in volume mixing ratio. For the transit spectra, the planetary radius, which here corresponds to the radius at 10 bar, was allowed to vary between $\pm50\%$ of the literature value. In emission, we set the planet radius to the best-fit value from our transmission retrieval.

The cloud top pressure prior ranged from $10^{-2}$ to $10^6$ Pa, in log-uniform scale. For the day side, we do not consider clouds in the model. In transmission, the temperature-pressure profile was assumed to be isothermal while in emission a 3-point profile was used. 

Finally, we use Multinest \citep{Feroz_Hobson_2008, Feroz_2009, Feroz_2013} with 1500 live points and an evidence tolerance of 0.5 in order to explore the likelihood space of atmospheric parameters.

\subsection{Modelling Equilibrium Chemistry - petitCODE}

To help contextualize our free retrieval results, we computed a self-consistent forward model with petitCODE, a 1D pressure-temperature iterator solving for radiative-convective and chemical equilibrium \citep{molliere_petitcode, molliere_jwst}. The code includes opacities for H$_2$, H$^-$, H$_2$O, CO, CO$_2$, CH$_4$, HCN, H$_2$S, NH$_3$, OH, C$_2$H$_2$, PH$_3$, SiO, FeH, Na, K, Fe, Fe$^+$, Mg, Mg$^+$, TiO and VO, as well as radiative scattering and collision induced absorption by H$_2$--H$_2$ and H$_2$--He. The atmosphere computed with petitCODE is assumed to be cloud-free, but the possibility of condensing refractory species is included in the equilibrium chemistry. Our petitCODE model for KELT-7 b was computed using the stellar and planetary parameters determined by \cite{Bieryla_2015}. Here, the surface gravity was computed using the planetary mass and radius. Furthermore, an intrinsic temperature of 600K was adopted, in accordance with its high equilibrium temperature \citep{thorngren_2019}. Finally, a global planetary averaged redistribution of the irradiation was assumed.

\subsection{Ephemeris Refinement}

Accurate knowledge of exoplanet transit times is fundamental for atmospheric studies. To ensure the KELT-7\,b can be observed in the future, we used our HST white light curve mid time, along with data from TESS \citep{ricker}, to update the ephemeris of the planet. TESS data is publicly available through the MAST archive and we use the pipeline from \cite{edwards_orbyts} to download, clean and fit the 2 minute cadence data. KELT-7\,b had been studied in Sector 19 and after excluding bad data, we recovered 9 transits. These were fitted individually with the planet-to-star radius ratio ($R_p/R_s$) and transit mid time ($T_{mid}$) as free parameters. We note that the ephemeris of KELT-7\,b was also recently refined by \cite{garhart_spitzer} and we also used the mid times derived in that study.\\

\begin{figure*}
    \centering
    \includegraphics[width = 0.95\textwidth]{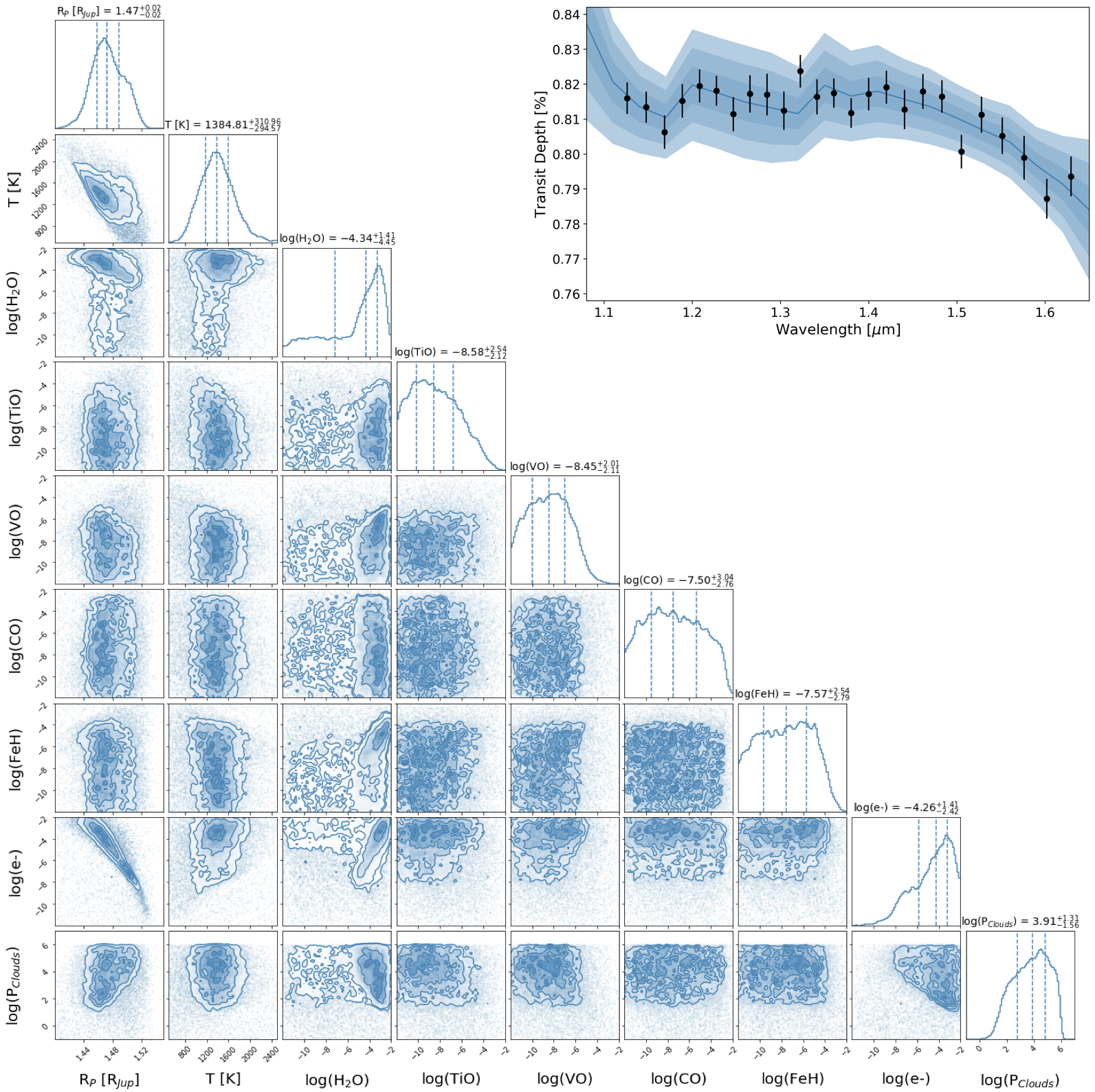}
    \caption{KELT-7\,b atmospheric retrieval posterior distributions of the transmission spectrum and the best-fit model with 3$\sigma$ confidence in blue while the HST WFC3 data is represented by the black data points.}
    \label{fig:transit_spec_post}
\end{figure*}{}

\begin{figure*}
    \centering
    \includegraphics[width = 0.95\textwidth]{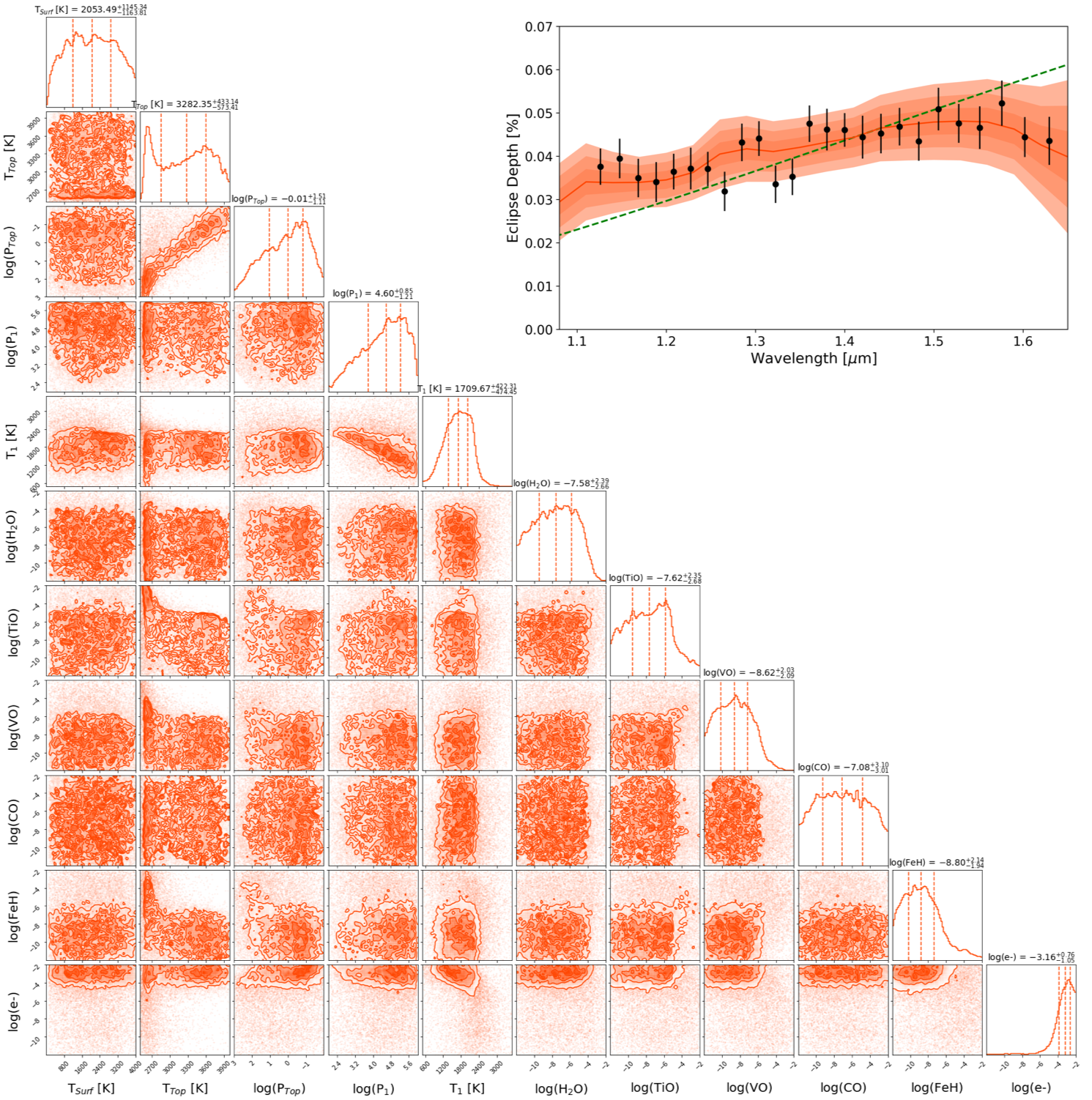}
    \caption{KELT-7\,b atmospheric retrieval posterior distributions of the emission spectrum and the best fit retrieved with 3$\sigma$ confidence in red while the HST WFC3 data is represented by the black data points. Also shown is a blackbody fit (green) which converged to a temperature of T$_{BB}$ $\simeq$ 2500 K and does not well describe the observations.}
    \label{fig:eclipse_spec_post}
\end{figure*}{}

\section{Results}
\label{results}

\begin{figure*}
    \centering
    \includegraphics[width = \columnwidth]{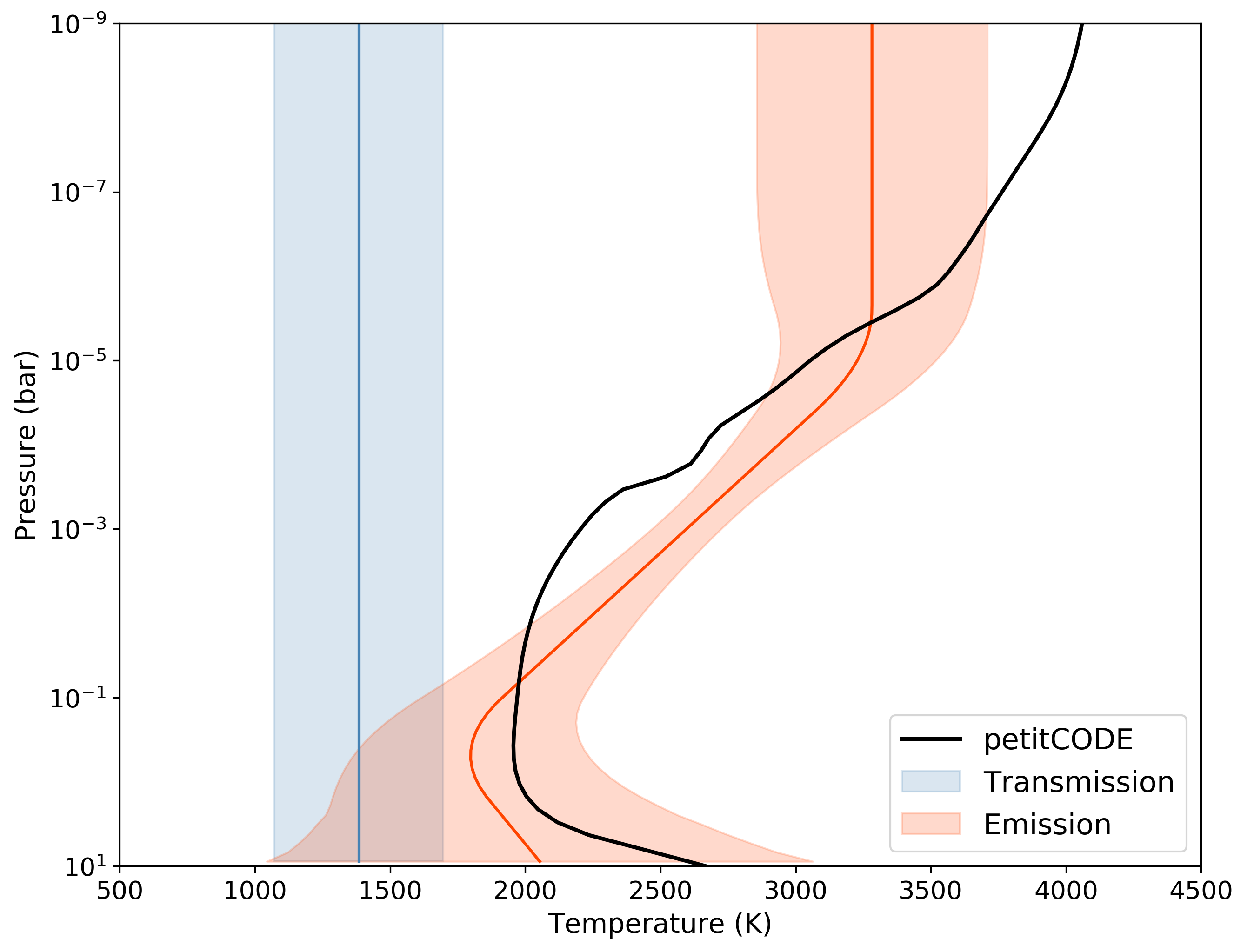}
    \includegraphics[width = \columnwidth]{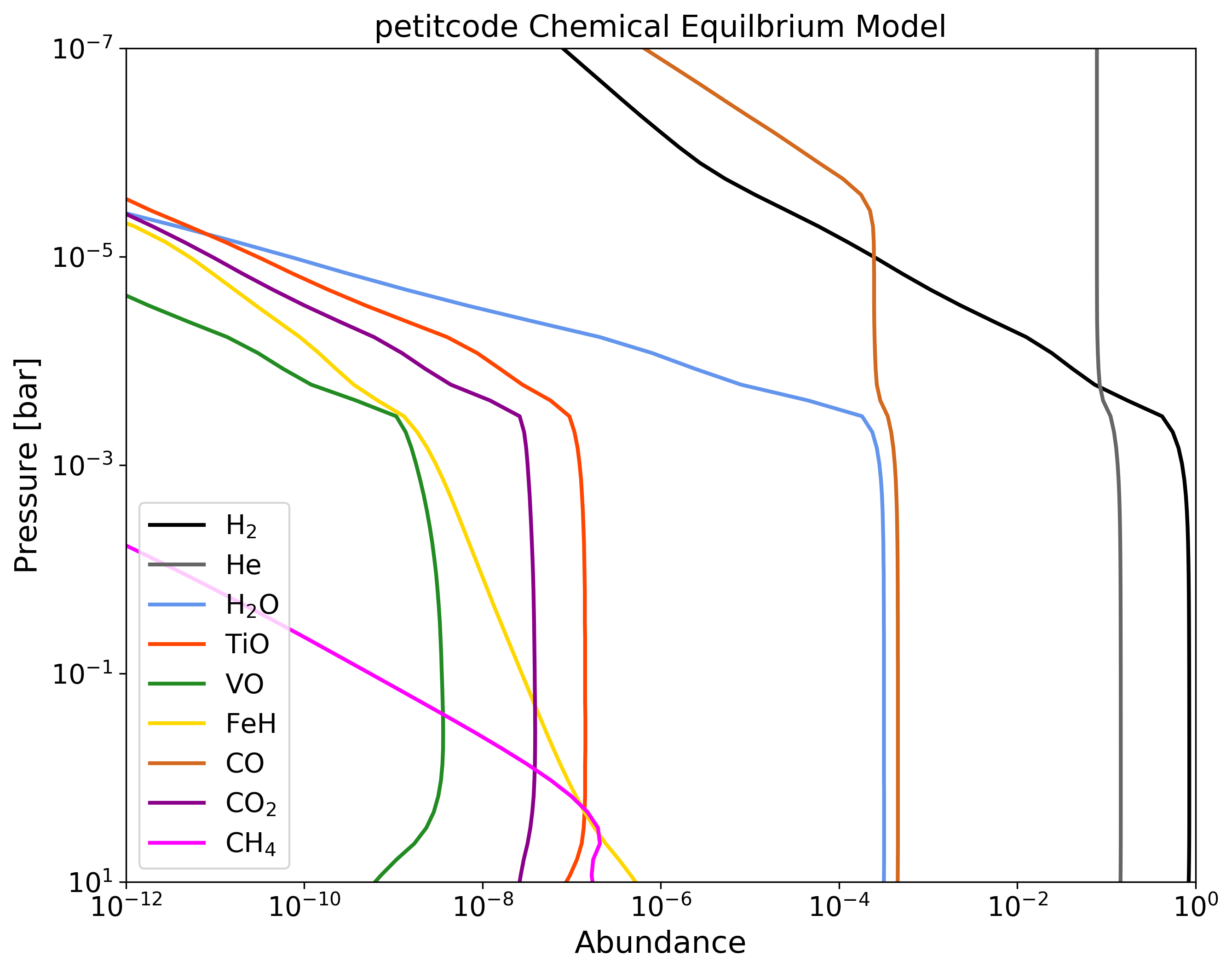}
    \includegraphics[width = \columnwidth]{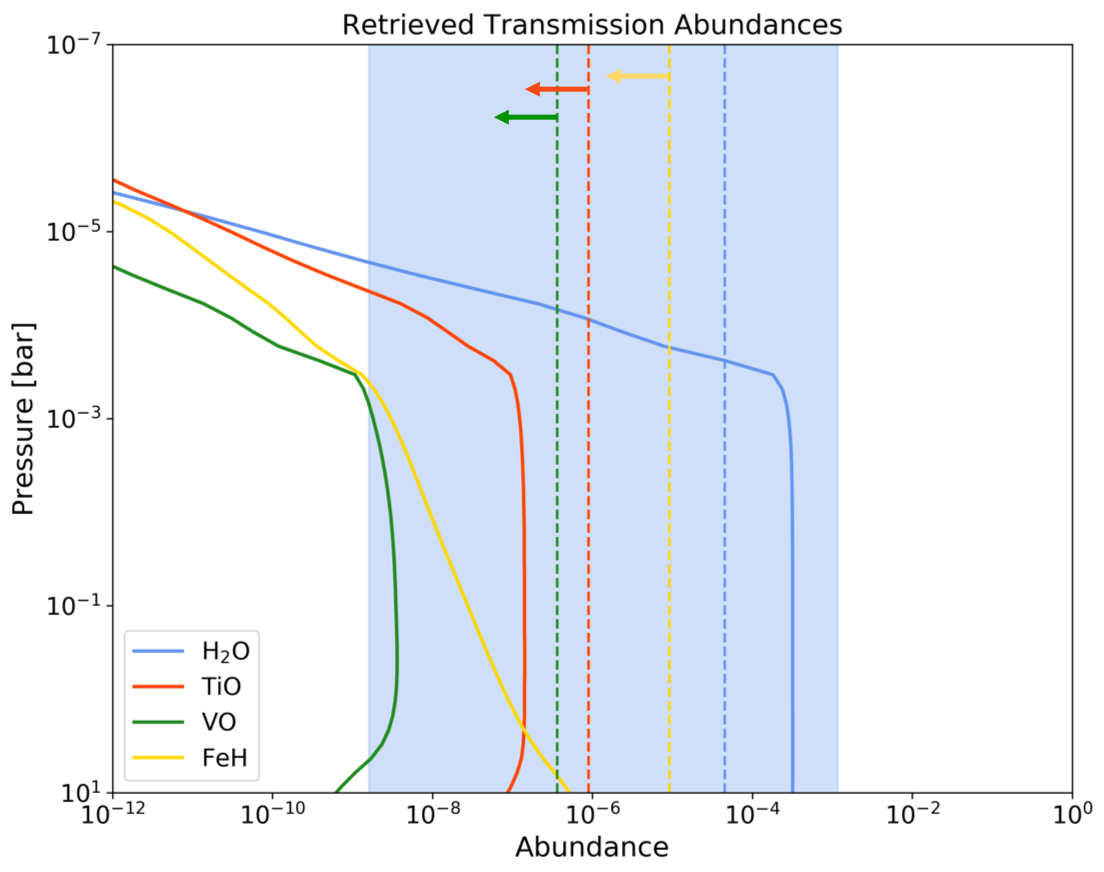}
    \includegraphics[width = \columnwidth]{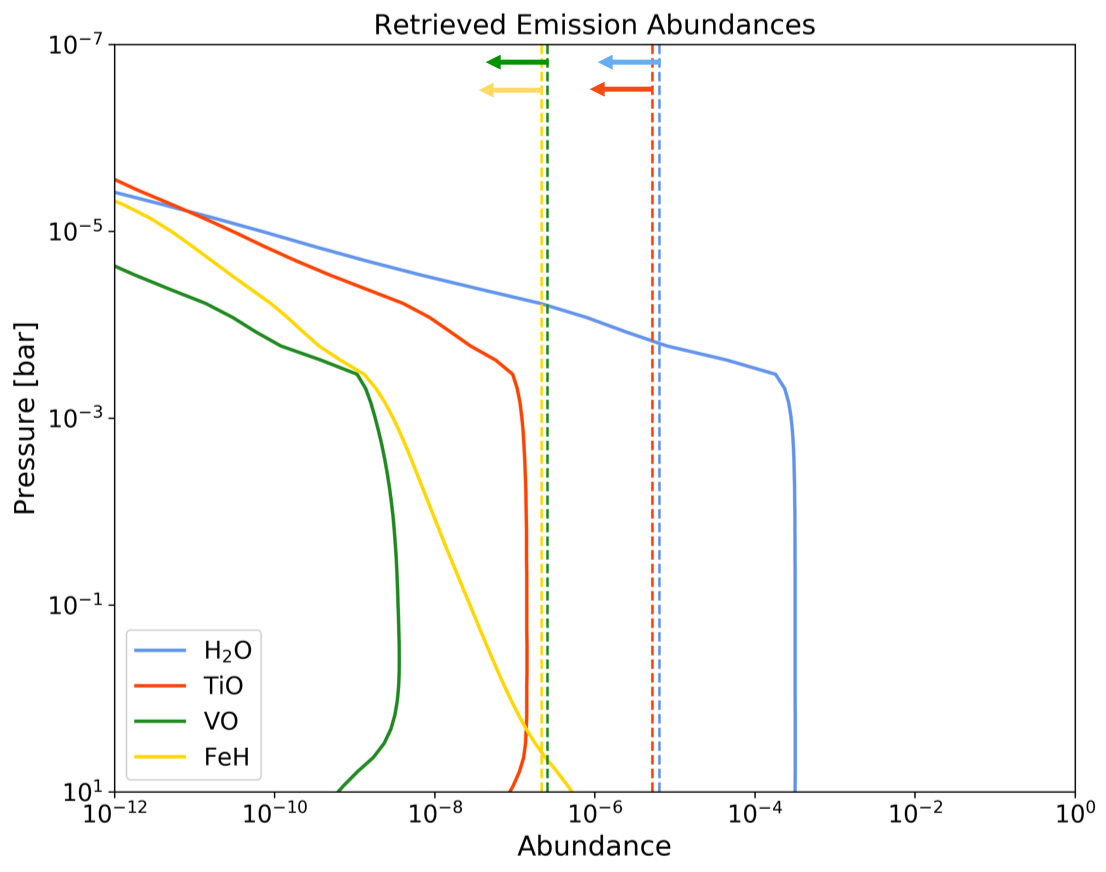}
    \caption{Results of our self-consistent petitCODE model for KELT-7 b and our retrievals on WFC3 data. Top left: comparison of the temperature-pressure profiles. The petitCODE model
    (black) features a thermal inversion at 1 mbar, because of absorption by TiO and VO, and closely matches the retrieved profile. Top right: Molecular abundances for the petitCODE simulation. The equilibrium fractions of most molecules remain approximately constant for pressures higher than 1 mbar, but drop quickly at lower atmospheric pressures due to thermal dissociation. Bottom left: Comparison of constrained molecular abundances in transmission (dotted lines) to those from the petitCODE simulation (solid lines). The water abundance is seen to be within 1$\sigma$ of that which is predicted. Bottom right: Comparison of upper bounds placed on molecular abundances in emission to those from the petitCODE simulation. The 1-$\sigma$ upper bound on the water abundance is significantly lower than the petitCODE simulations. In both transmission and emission, the 1-$\sigma$ upper bounds on TiO, VO and FeH are well above the predicted abundances, suggesting the data is not sufficient to comment on their presence/absence.}
    \label{fig:petitcode}
\end{figure*}

Our analysis uncovers a rich transmission spectrum which is consistent with a cloud-free atmosphere and suggests the presence of water and dissociated hydrogen, as shown by the posterior distributions in Figure \ref{fig:transit_spec_post}. We calculate the Atmospheric Detectability Index (ADI) \cite{angelos30} to be 16.8 for the transmission spectrum, indicating strong evidence of atmospheric features. The retrieved temperature of around 1400 K for the terminator region is consistent with the expected value given the equilibrium temperature. However, as we are in temperature regime (T$_{eq}$ $\simeq$ 2000 K) where 3D effects across the limb could occur, we may be biased on the abundances and temperature retrieved \citep{Pluriel_2020}.

In contrast, the extracted emission spectrum does not contain strong absorption or emission features. Although the data is not consistent with a simple blackbody, as shown in Figure \ref{fig:eclipse_spec_post}, it can be explained by a varying temperature-pressure profile and H-. We calculate the ADI against a simple blackbody to be 19.6, highlighting the poorness of the blackbody fit and indicating the presence of a detectable atmosphere. However, it could also indicate that 2D effects, such as those suggested by \citep{Wilkins2014}, are affecting the observed spectrum. In their study of CoRoT-2\,b they observed an emission spectrum similar to that of KELT-7\,b (i.e. one that is poorly fit by a blackbody but which can be better fit using 2 blackbodies). Such inhomogeneities will certainly be important in the analysis of emission data from future missions \citep{taylor_emission}. Our best fit favours a thermal inversion in the stratosphere of KELT-7\,b. Although the lower part of the atmosphere has large temperature uncertainties (between 1000 K and 3000 K), the middle and top temperature-pressure points are well-constrained enough to indicate a thermal inversion with a temperature at the top between 2900 K and 3700 K as shown in Figure \ref{fig:petitcode}. Both transmission and emission spectra, along with their best-fit solutions are shown in Figures \ref{fig:transit_spec_post} and \ref{fig:eclipse_spec_post}.

The retrieved emission temperature-pressure profile closely agrees with the petitCODE simulations, both showing a temperature inversion as shown in Figure \ref{fig:petitcode}. The retrieved water abundance in transmission is consistent with the predictions although the 1$\sigma$ upper bound in emission is below what is expected. For TiO, VO and FeH, the upper bound place on their abundances is significantly above the amount expected from our petitCODE simulations, suggesting the HST WFC3 data is no sensitive enough for us  to conclude on their presence/absence.

\begin{table}
    \centering
    \resizebox{\columnwidth}{!}{
    \begin{tabular}{cccc}
    \hline \hline
    Retrieved parameters  & Bounds & Transmission & Emission \\ \hline\hline
    $\log[H_2O]$  & -12 - -1  & $-4.34^{+1.41}_{-4.45}$ & $< -5$ \\
    $\log[e^{-}]$  & -12 - -1  & $-4.26^{+1.41}_{-2.42}$ & $< -5$  \\
    $\log[FeH]$  & -12 - -1  & $< -5$ & $< -5$  \\
    $\log[TiO]$  & -12 - -1  & $< -5$ & $< -5$ \\
    $\log[VO]$  & -12 - -1  & $< -5$ & $< -5$\\
    $\log[CO]$  & -12 - -1  & $< -5$ & $< -5$ \\
    \hline
    $\mu$ (derived) & - & $ 2.33^{+0.28}_{-0.02}$ &  $2.31^{+0.22}_{-0.00}$  \\
    $R_p\, (R_{jup})$ & $\pm $ 50\% & $1.47^{+0.02}_{-0.02}$ & - \\ 
    $\log{P_{clouds}}$ & -2 - 6 & $3.91^{+1.31}_{-1.56}$ & -  \\
    \hline
    $T_p$ [K] & 1000-4000 & $ 1385^{+311}_{-295}$ & - \\
    \hline
    $T^{bot}_{p}$ [K] & 1000-4000 & - & $ 2053^{+1145}_{-1164}$ \\
    $T^{mid}_{p}$ [K] & 1000-4000 & - & $ 1710^{+422}_{-474}$ \\
    $T^{top}_{p}$ [K] & 1000-4000 & - & $ 3282^{+463}_{-573}$ \\
    \hline
    ADI & - & 16.8 & 19.6 \\ 
    $\sigma$-level & - & $>5\sigma$ & $>5\sigma$ \\
    \hline
    \multicolumn{4}{c}{Updated Ephemeris}\\\hline
\multicolumn{2}{c}{Period [days]} & \multicolumn{2}{c}{2.73476541$\pm$0.00000036} \\
\multicolumn{2}{c}{T$_0$ [BJD$_{TDB}$]} & \multicolumn{2}{c}{2458384.425577$\pm$0.000099} \\\hline\hline
    \end{tabular}
    }
    \caption{Table of fitted parameters for the retrievals of KELT-7\,b transmission and emission spectra for HST data only and the updated ephemeris for the planet.}
    \label{tab:retrieval}
\end{table}

\begin{figure}
    \centering
    \includegraphics[width = \columnwidth]{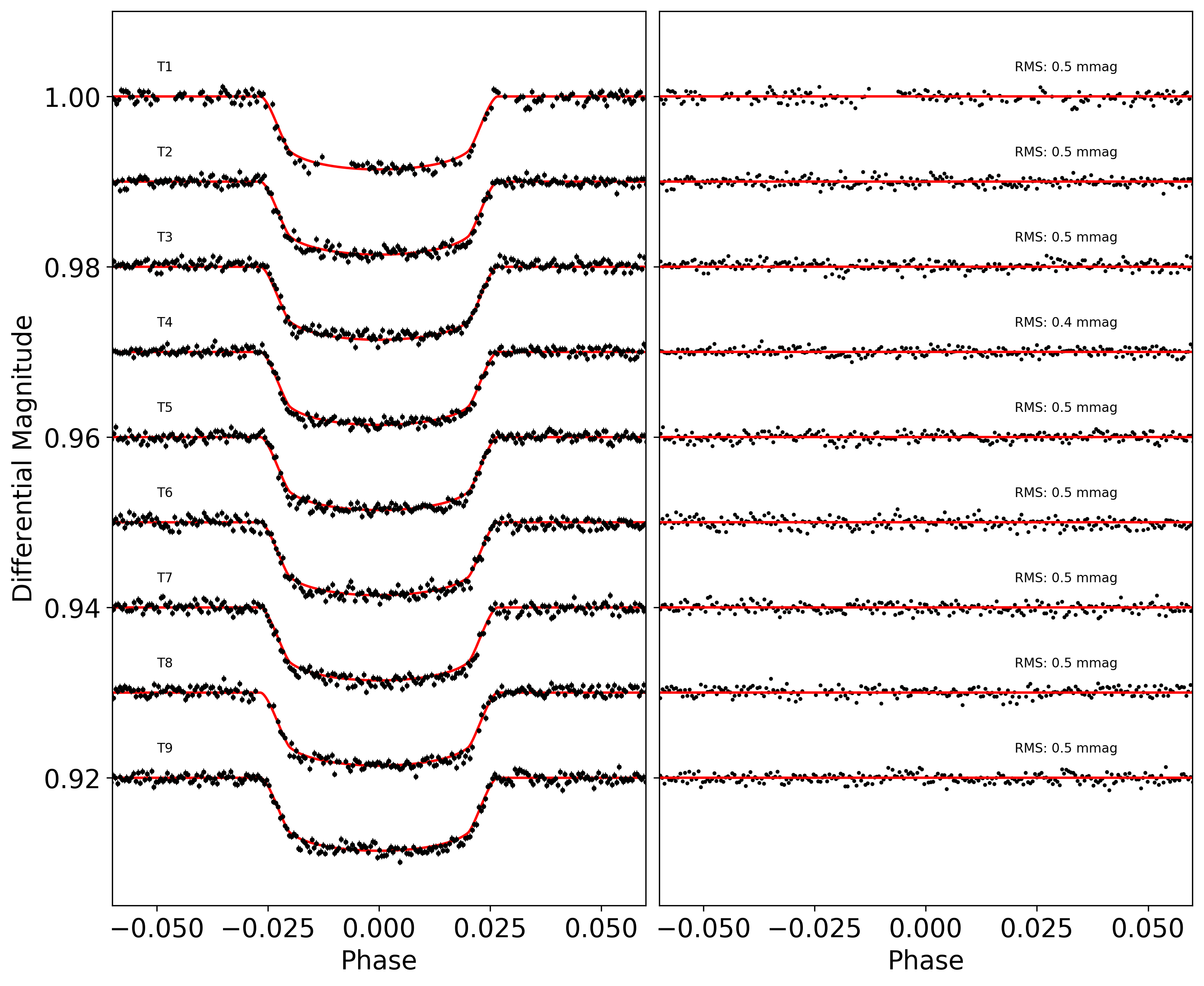}
    \includegraphics[width = \columnwidth]{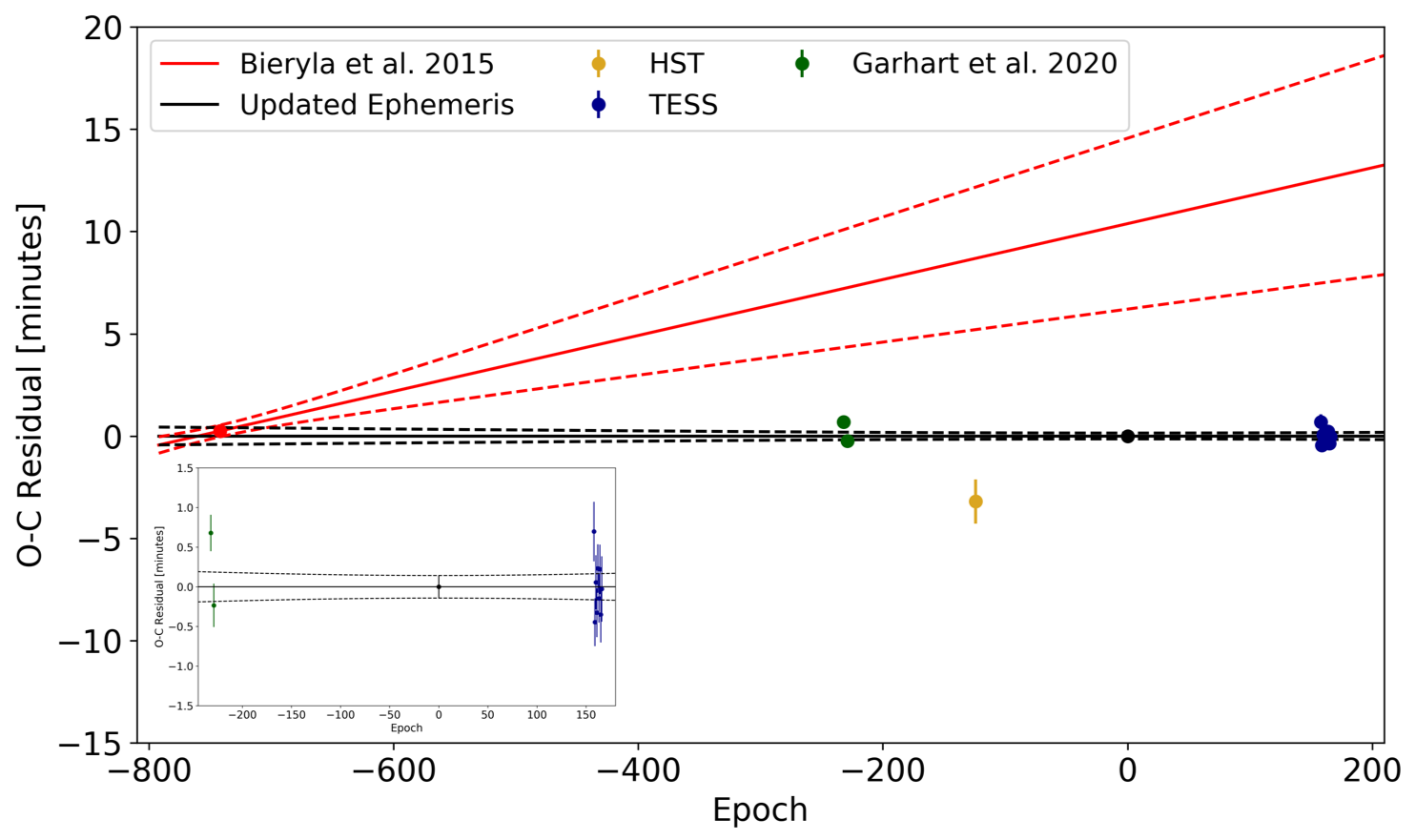}
    \caption{Top: TESS observations of KELT-7\,b presented in this work. Left: detrended data and best-fit model. Right: residuals from fitting. Bottom: Observed minus calculated (O-C) mid-transit times for KELT-7\,b. Transit mid time measurements from this work are shown in gold (HST) and blue (TESS), while the literature T$_0$ value is in red. The black line denotes the new ephemeris of this work with the dashed lines showing the associated 1$\sigma$ uncertainties and the black plot data point indicating the updated T$_0$. For comparison, the previous literature ephemeris and their 1$\sigma$ uncertainties are given in red. The inset figure shows a zoomed plot which highlights the precision of the TESS mid time fits.}
    \label{fig:k7_ephm}
\end{figure}

The transits of KELT-7\,b from HST and TESS were seen to arrive early compared to the predictions from \cite{Bieryla_2015}. As the observation did not include ingress or egress, the HST transit fitting is not as precise, and potentially not as accurate, as the TESS data which captured the whole event. Using this new data, we determined the ephemeris of KELT-7\,b to be P = 2.73476541$\pm$0.00000036 days and T$_0$ = 2458384.425577$\pm$0.000099 BJD$_{TDB}$ where P is the planet's period, T$_0$ is the reference mid-time of the transit and BJD$_{TDB}$ is the barycentric Julian date in the barycentric dynamical. Our derived period is 0.84 s and 0.063 s shorter than the periods from \cite{Bieryla_2015} and \cite{garhart_spitzer} respectively. The $\sim$10 minute residual of the TESS transits from the ephemeris of \cite{Bieryla_2015} show the necessity to regularly follow-up transiting planets and for programmes such as ExoClock\footnote{\url{https://www.exoclock.space}} to which our observations have been uploaded. By the launch of Ariel in late 2028, around 1400 orbits of KELT-7\,b will have occurred since the T$_0$ derived here and the difference in the predicted transit times between the ephemeris from this work and from \cite{garhart_spitzer} would be around 5 minutes. The observed minus calculated residuals, along with the fitted TESS light-curves are shown in Figure \ref{fig:k7_ephm} while the fitted mid times can be found in Table \ref{tab:mid_times}.

\begin{table}
    \centering
    \resizebox{\columnwidth}{!}{
    \begin{tabular}{ccc} \hline\hline
     Epoch & Transit Mid Time [BJD$_{TDB}$] & Reference  \\\hline\hline
    -742 & 2456355.229809 $\pm$ 0.000198 & \cite{Bieryla_2015} \\
    -232 & 2457749.959530 $\pm$ 0.000160 &  \cite{garhart_spitzer} \\
    -229 & 2457758.164460 $\pm$ 0.000190 &  \cite{garhart_spitzer} \\
    -124 & 2458045.316888 $\pm$ 0.000627 &  This Work* \\
     158 & 2458816.518025 $\pm$ 0.000282 &  This Work \\
     159 & 2458819.253548 $\pm$ 0.000220 &  This Work \\
     160 & 2458821.987982 $\pm$ 0.000202 &  This Work \\
     161 & 2458824.723075 $\pm$ 0.000183 &  This Work \\
     162 & 2458827.457521 $\pm$ 0.000214 &  This Work \\
     163 & 2458830.192334 $\pm$ 0.000196 &  This Work \\
     164 & 2458832.927100 $\pm$ 0.000212 &  This Work \\
     165 & 2458835.661872 $\pm$ 0.000246 &  This Work \\
     166 & 2458838.396646 $\pm$ 0.000194 &  This Work \\ \hline\hline
  \multicolumn{3}{c}{*Data from Hubble}\\\hline \hline
    \end{tabular}
    }
    \caption{Transit mid times used to refine the ephemeris of planets from this study. All mid times reported in this work are from TESS unless otherwise stated.}
    \label{tab:mid_times}
\end{table}

\section{Discussion}

In transmission, H$_2$O and H$^{-}$, ($\mathrm{log}(\mathrm{H_2O})=-4.34^{+1.41}_{-4.45}$ and $\mathrm{log}(\mathrm{H^{-}})=-4.26^{+1.41}_{-2.42}$) are well defined. As shown in the posterior distributions in Figure \ref{fig:transit_spec_post}, correlations exist between the abundances of the molecules, particularly between H$_2$O and H$^{-}$. We could expect to also find CO in such a hot atmosphere ($\mathrm{T}_{eq}\,\simeq\,2000 \mathrm{K}$), evidence for which has been seen for other hot Jupiters: WASP-121b \citep{Evans_2017, Parmentier_2018} and WASP-33b \citep{Haynes_2015}. However our retrieval analysis on HST data only provides no evidence for the presence of these molecules. We also explored the addition of Spitzer/IRAC and TESS data, see Section \ref{Spitzer-analysis} for more details.  

The non-detection of TiO and VO in the terminator region agreed with predictions from \cite{Spiegel_2009_can_tio}. Their work suggests that in highly irradiated atmospheres, similar to KELT-7\,b, TiO and VO would be likely to rain out in cold traps and disappear from the visible atmosphere. Observational evidence for these processes was reported in emission spectrum of Kepler-13Ab \cite{beatty_kep13}. To overcome these effects, large advective mixing and higher temperatures (higher than 1800K) would be required, leading to large abundance of VO in the day side of the planet. 
However, in our analysis, we do not find evidence for VO or TiO on the day side of KELT-7\,b, which could be unravelled with higher signal-to-noise in future observations such as Ariel or JWST. Also, our analysis potentially finds a large difference in the day/night temperatures with thermal inversion on the dayside, which would indicate that day and night side of the planet are poorly coupled by large scale dynamical processes, thus preventing VO and TiO from reaching the cold night side and condensating. \cite{Fortney_2008} postulated that the presence of optical absorbers would lead, and require, large day - night temperature contrasts, which could suggest that TiO and VO are present but simply beyond the sensitivity of the data. 


\subsection{Addition of Spitzer data}
\label{Spitzer-analysis}

KELT-7\,b has also been studied, in both transmission and emission, by the Spitzer Space Telescope using the each of the 3.6 $\mu m$ and 4.5 $\mu m$ channels of the InfraRed Array Camera (IRAC). Combining data from multiple instruments or observatories has become the standard within the field as a way of increasing the spectral coverage, seeking to break the degeneracies that occur when fitting data over a narrow wavelength range \citep[e.g.][]{sing}. However, such a procedure is fraught with risk due to potential incompatibilities between datasets. Firstly, it has been shown using different orbital parameters (a/Rs and i) in the fitting of the data can lead to offsets between the datasets \citep{yip}. Secondly, the choice of limb darkening coefficients can cause vertical shifts in the spectrum \citep{angelos30}. Thirdly, stellar variability, activity or spots can also induce offsets in the recovered spectra \citep[e.g.][]{wasp52b}. Finally, imperfect correction of instrument systematics can alter the fitted white light curve depth, again generating shifts between datasets, making the derived transit/eclipse depths incompatible \cite[e.g.][]{stevenson_wasp12, stevenson_gj436, diamond_hd209}. Each of these effects can of course affect HST data alone but, in this case, the offset would likely only lead to slight changes in the retrieved temperature or radius. When combining datasets, for instance HST WFC3 and Spitzer IRAC, offsets in one of these, or differing offsets in both, could lead to wrongly recovered abundances. In transmission and emission, Spitzer observations are sensitive to CH$_4$ in the 3.6$\mu$m band and CO or CO$_2$ in the 4.5$\mu$m band while WFC3 data cannot constrain these molecules. Thus the detection, or non-detection, would be based entirely off the two Spitzer bands relative to the WFC3 data. An offset in either one of these would instigate the incorrect recovery of the CH$_4$, CO and CO$_2$ abundances. In emission, these Spitzer points have been used as evidence for the presence of, or lack of, a thermal inversion. Differences in the correction of systematics have led to discrepant results (e.g. HD\,209458\,b: \citet{knutson_hd209, diamond_hd209}).

Here, we tentatively add the Spitzer data for KELT-7\,b, taken from the study by \cite{garhart_spitzer} of 36 hot Jupiters. For the fitting of the Spitzer eclipses, \cite{garhart_spitzer} froze the orbital parameters to those from \cite{Bieryla_2015}, overcoming the first hurdle about combining datasets. The transit observations were also fitted with fixed orbital parameters and limb darkening coefficients from \cite{Claret}, again showing consistency with our data. The latter two issues, of stellar variability or activity and the detrending of instrument systematics, cannot be easily determined without an overlap in spectral coverage. Therefore, we caution the reader that the compatibility of the datasets cannot be guaranteed. Additionally, we cautiously added the TESS data which was again fitted with same orbital parameters and limb darkening laws.

\begin{figure*}[]
    \centering
    \includegraphics[width = 0.45\textwidth]{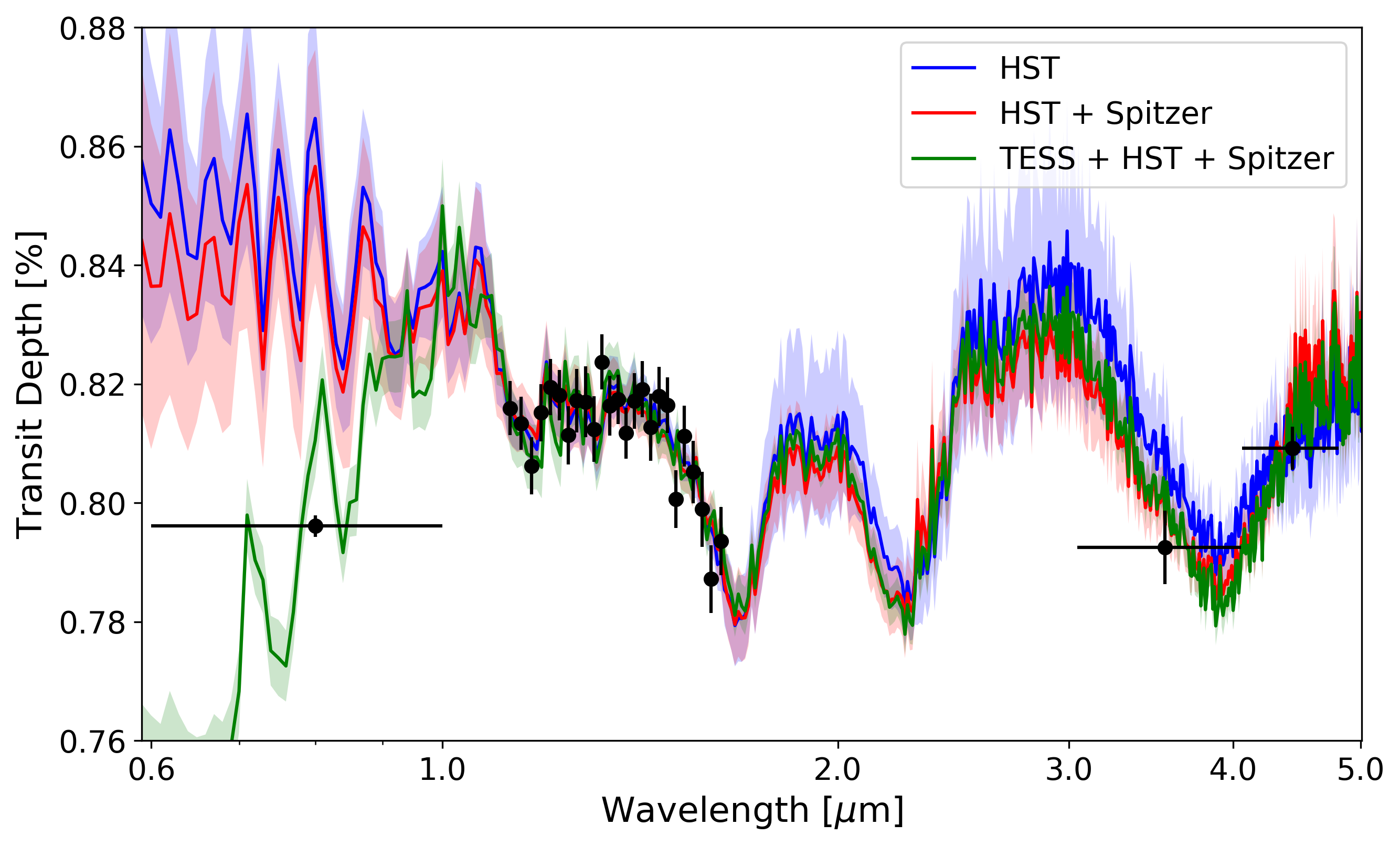}
    \includegraphics[width = 0.45\textwidth]{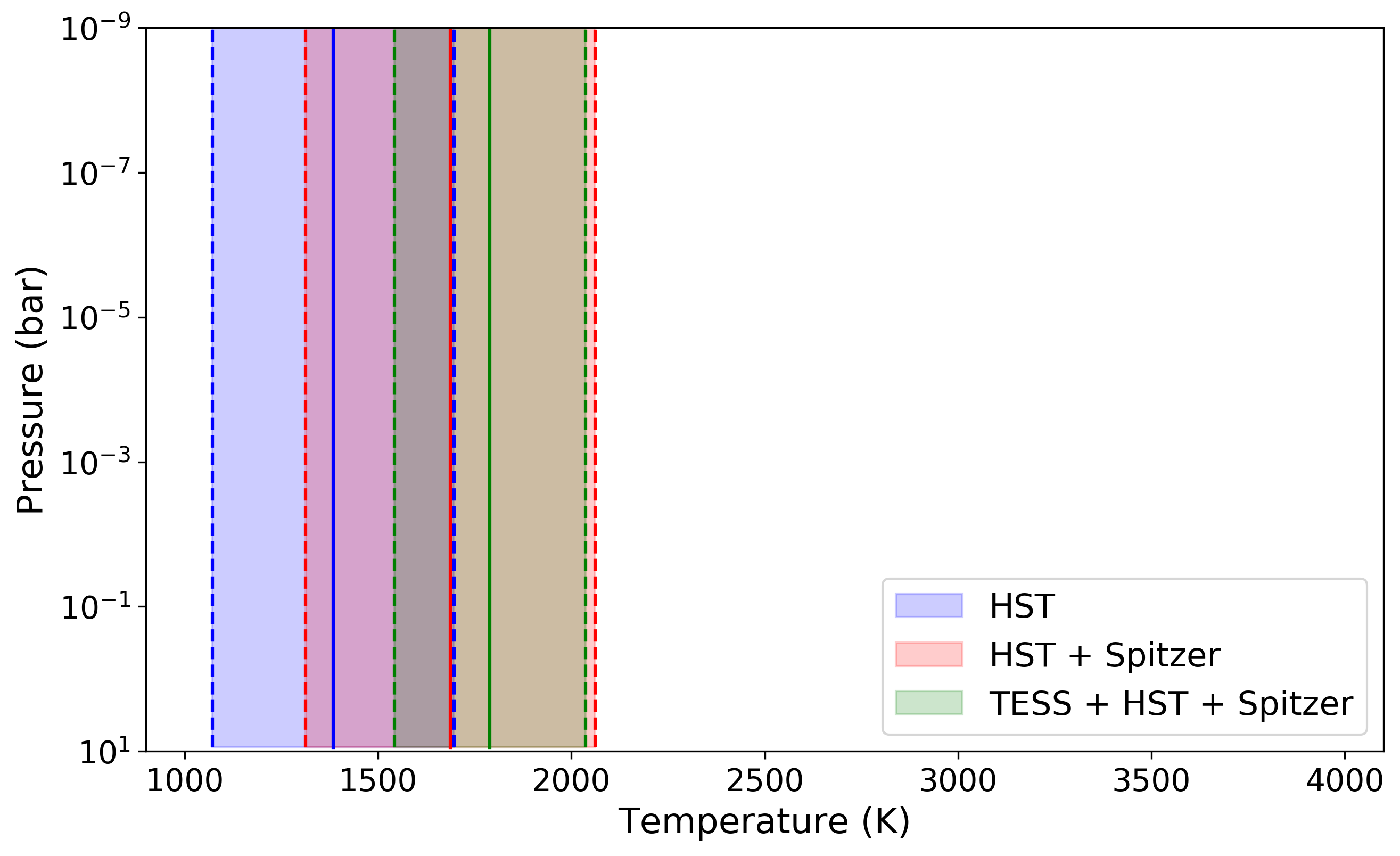}
    \includegraphics[width = 0.45\textwidth]{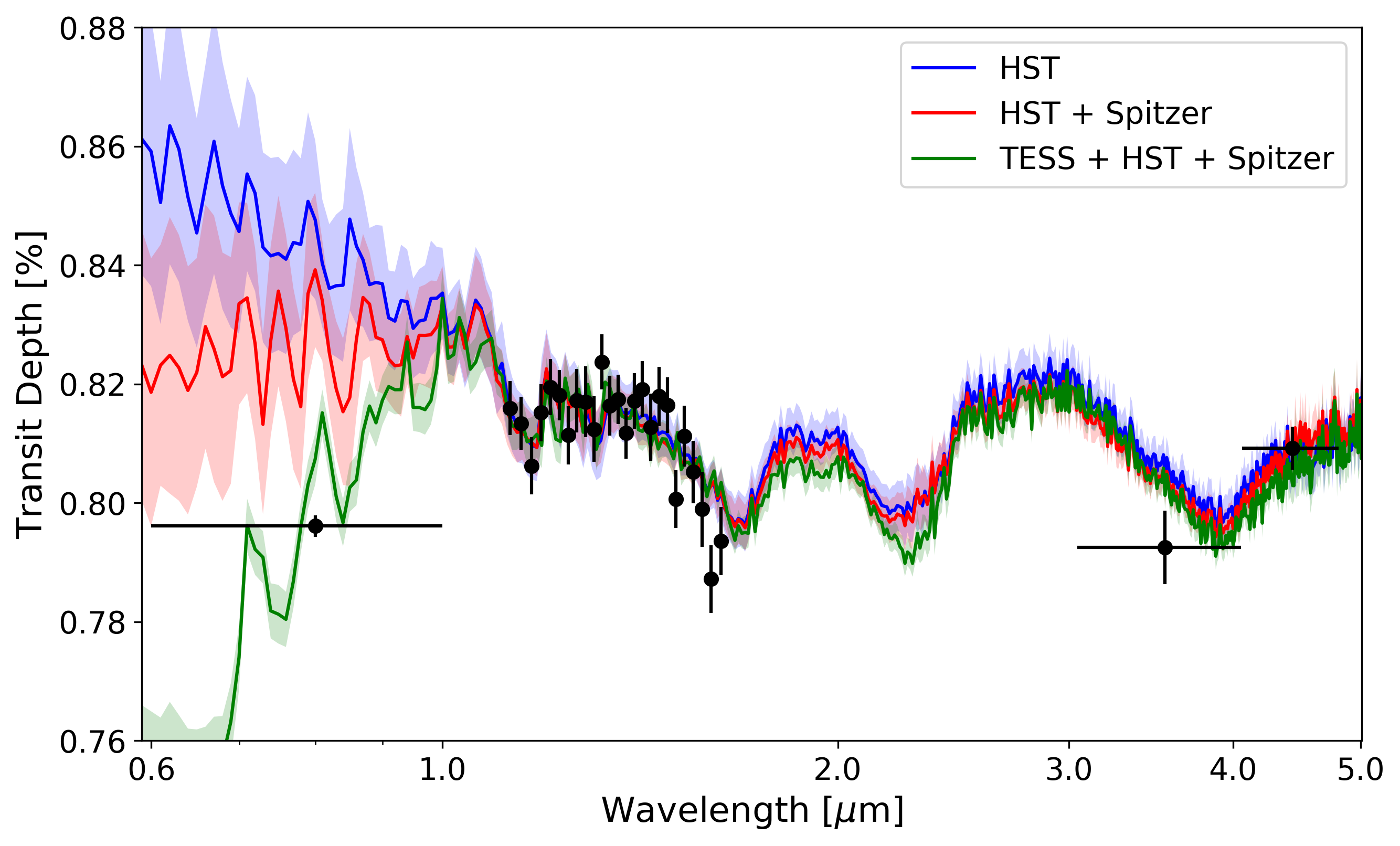}
    \includegraphics[width = 0.45\textwidth]{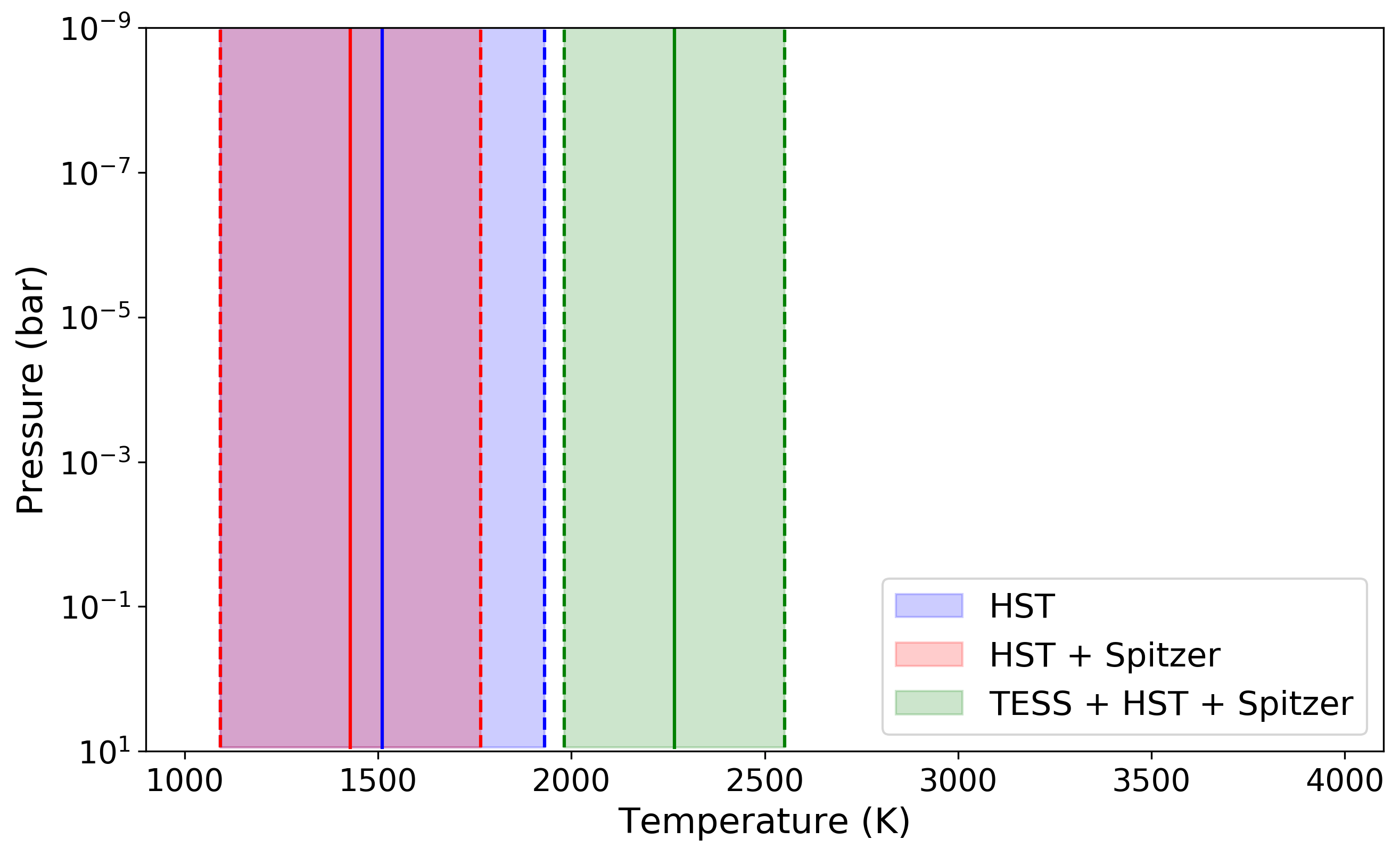}
    \includegraphics[width = 0.45\textwidth]{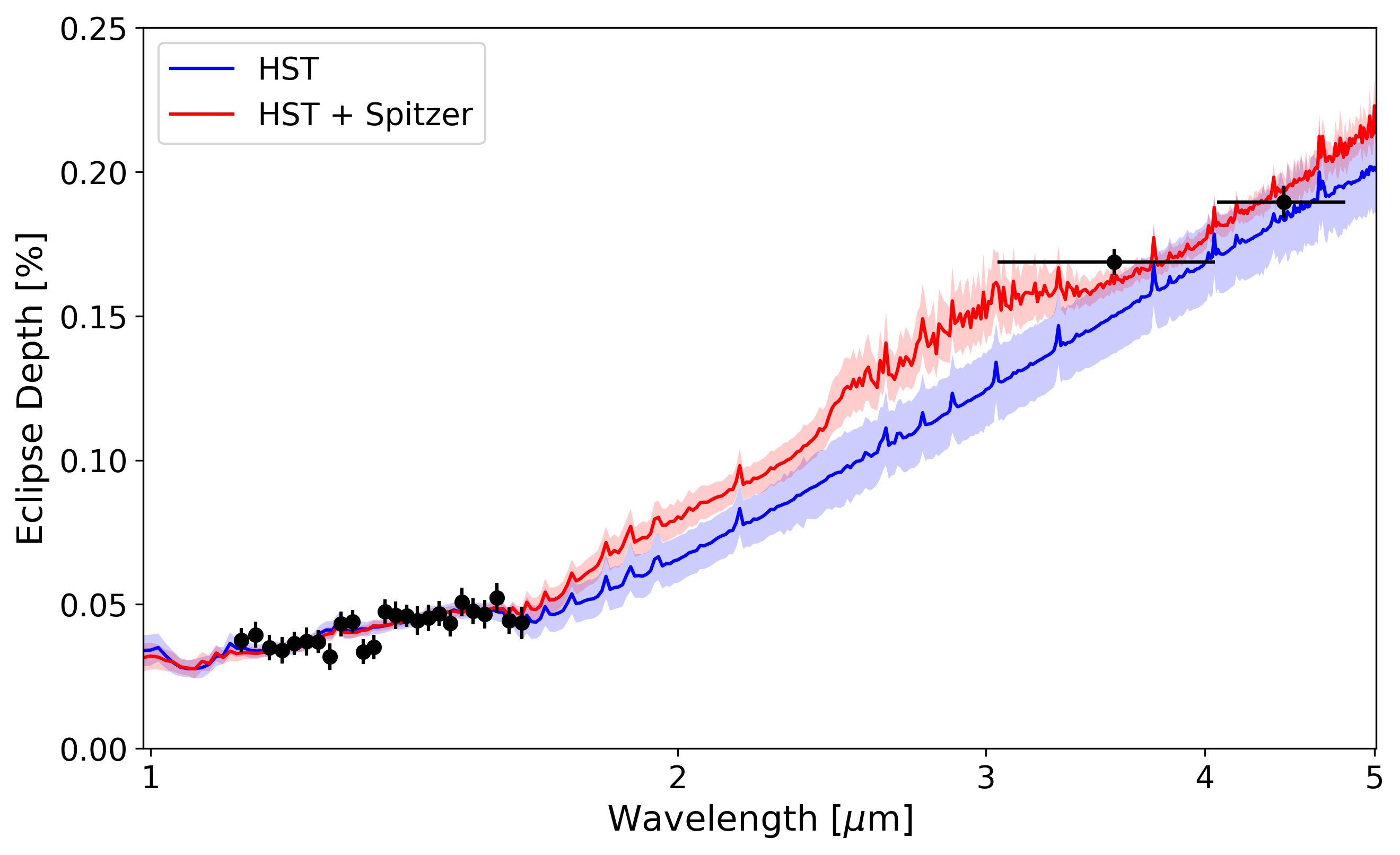}
    \includegraphics[width = 0.45\textwidth]{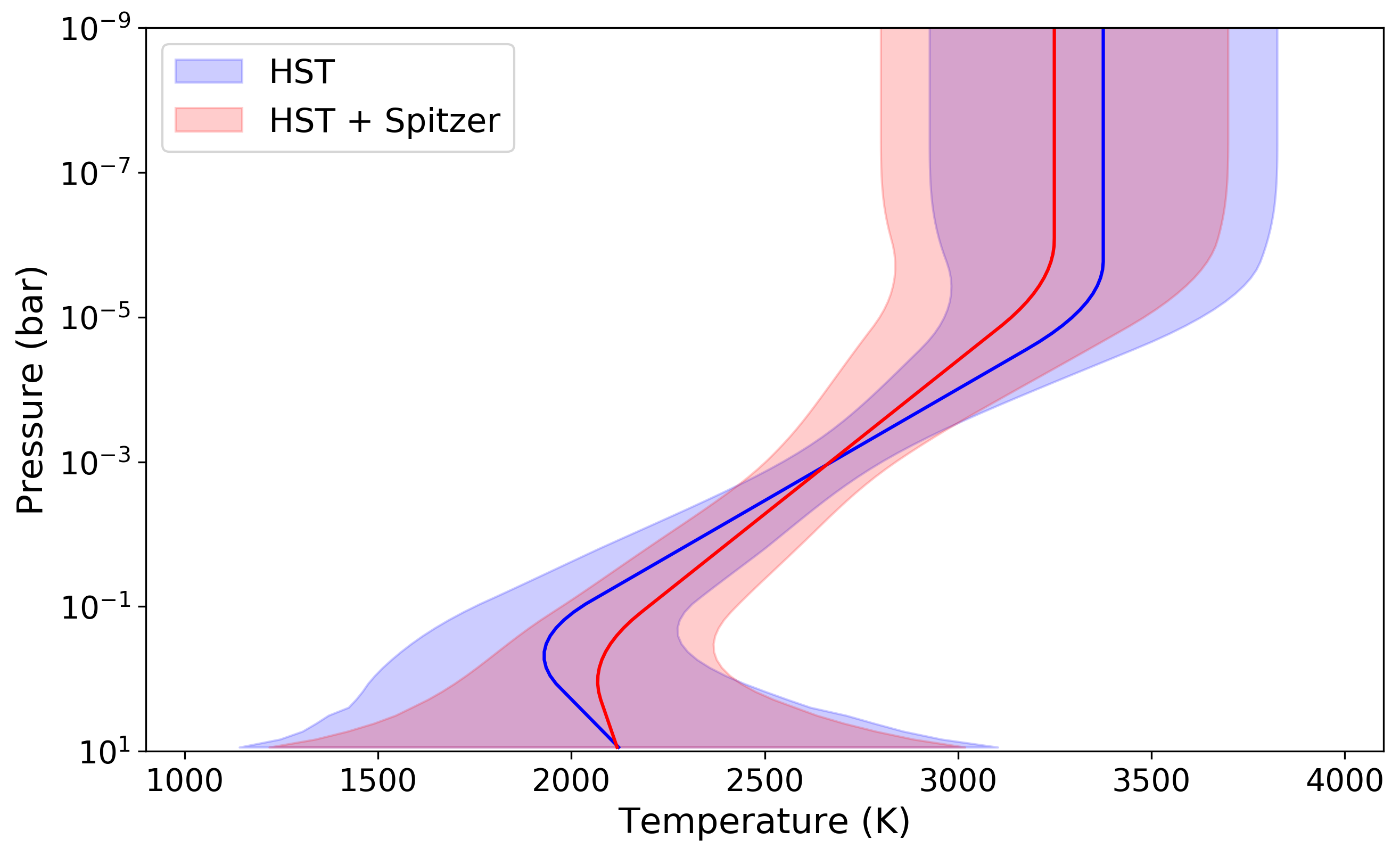}
    \caption{Best-fit spectra (left) and temperature-pressure profiles (right), with one sigma errors in each case, for transit (top and middle) and eclipse (bottom) observations of KELT-7\,b with HST data only, HST and Spitzer (red) and HST, Spitzer and TESS (green). The top transmission plots, and the emission case, include the H- opacity while the middle plots do not.}
    \label{fig:spitzer_results}
\end{figure*}

The best-fit retrieved spectra, both with and without the Spitzer data, is shown in Figure \ref{fig:spitzer_results}. Little difference is seen between the fits, particularly in transmission where the best-fits are almost constantly within 1$\sigma$ of each other. The recovered temperature pressure profiles are also practically identical. While this may suggest the data is compatible, it also shows that the information content of adding Spitzer, in this case, is relatively low. Therefore this begs the question of whether risking data incompatibility is worthwhile when there is little to gain.  Figure \ref{fig:posteriors_trans} and \ref{fig:posteriors_ec} show posteriors from the fittings with and without Spitzer, again highlighting the similarity between the fits. In transmission the only noticeable difference is in the recovered CO abundance, which is not constrained in the case of HST alone but the addition of Spitzer suggests an abundance of $log(CO) = -4.56^{+1.72}_{-4.69}$. The second change is in the water abundance recovered in emission, with HST and Spitzer converging to $log(H_2O) = -5.11^{+1.37}_{-4.03}$ while no constraint can be made in the HST only case.

On the other hand, the addition of the TESS transmission data drastically changes the solution, removing the detection of dissociated hydrogen, instead preferring FeH to explain the absorption at the shorter wavelengths within the G141 grism due to the shallow TESS transit depth. We also explored the retrieved atmospheric abundances in transmission without the H- opacity, again finding that there is little difference when adding Spitzer data as shown in Figure \ref{fig:posteriors_nohminus}. In this case, all data combinations readily agree on the abundances of H$_2$O, FeH and CO. However, differences in seen in the evidence for TiO and VO, the presence of which the addition TESS data of rules out for log(TiO,VO) $>$ -10, and in the recovered radius and terminator temperature, with the TESS dataset preferring a lower radius and higher temperature. We suggest it is imperative to, at the very least, study multi-instrument data sets separately, as well as combined, when doing model fitting, as we have done in this study.

\subsection{Future Characterisation}

The most effective solution to understanding the source of the absorption seen in transmission between 1.1-1.3 $\mu m$ would be to take more data, namely with the G102 grism of WFC3 which covers 0.8-1.1 $\mu$m. As there is also archival STIS transmission data for KELT-7\,b, with the G430L and G750L grisms, this would provide continuous coverage from 0.3-1.6 $\mu$m, allowing for better constraints on the abundances of these optical absorbers. We would also advocate for additional eclipse observations with WFC3, with either grism, to increase the spectral coverage and/or increase the signal to noise, which may allow for spectral features to be uncovered.

Additionally, future space telescopes JWST \citep{greene}, Twinkle \citep{twinkle} and Ariel \citep{tinetti_ariel} will provide a far wider wavelength range and these missions will definitively move the exoplanet field from an era of detection into one of characterisation, allowing for the identification of the molecular species present and their chemical profile, insights into the atmospheric temperature profile and the detection and characterisation of clouds. Ariel, the ESA M4 mission due for launch in 2028, will conduct a survey of $\sim$1000 planets to answer the question: how chemically diverse are the atmospheres of exoplanets? KELT-7\,b has been identified as an excellent target for study with Ariel \citep{ariel_tl}, through both transmission and emission spectroscopy, and simulated error bars from \citet{mugnai} have been added to the best-fit spectra to showcase this. Figure \ref{fig:simulations} shows simulated Ariel and JWST observations and highlights the wavelength coverage performed by those future missions. Additionally ExoWebb \citep{edwards_exowebb} has been used to showcase the capability of JWST for studying this planet.

\section{Conclusion}

We present spectroscopic transmission and emission observations of KELT-7\,b taken with Hubble WFC3. While the transit spectra demonstrates strong absorption features indicative of H$_2$O and H$^{-}$, the emission spectrum lacks features and can be fitted with CIA alone. We also explore adding data from Spitzer IRAC, with the results being very similar in both transmission and emission. Finally, we find that adding TESS data in our analysis strongly modifies our results. As these instruments do not provide spectral overlap, more data is needed to fully understand the source of the optical absorption seen in transmission. Further observations with Hubble, or with the next generation of observatories, will undoubtedly allow for an enhanced probing of the atmosphere of this intriguing planet. The analysis of archival Hubble data is an essential preparatory step in enriching our comprehension of exoplanet atmospheres, allowing us to begin to appreciate their true diversity and understand the optimal observation strategy for upcoming facilities.
\\
\\
\\
\textbf{Acknowledgements:} This work was realised as part of ARES, the Ariel Retrieval Exoplanet School, in Biarritz in 2019. The school was organised by Jean-Philippe Beaulieu, Angelos Tsiaras and Ingo Waldmann with the financial support of CNES. 

This work is based upon observations with the NASA/ESA Hubble Space Telescope, obtained at the Space Telescope Science Institute (STScI) operated by AURA, Inc. The publicly available HST observations presented here were taken as part of proposal 14767, led by David Sing \citep{sing_proposal}. These were obtained from the Hubble Archive which is part of the Mikulski Archive for Space Telescopes. This paper includes data collected by the TESS mission which is funded by the NASA Explorer Program. TESS data is also publicly available via the Mikulski Archive for Space Telescopes (MAST). We are thankful to those who operate this archive, the public nature of which increases scientific productivity and accessibility \citep{peek2019}. This work is also based in part on observations made with the Spitzer Space Telescope, which is operated by the Jet Propulsion Laboratory, California Institute of Technology, under a contract with NASA. Finally, we thank our anonymous referee for suggesting we include the Spitzer data which, along with other useful comments, led to the improvement of the manuscript.

JPB acknowledges the support of the University of Tasmania through the UTAS Foundation and the endowed Warren Chair in Astronomy, Rodolphe Cledassou, Pascale Danto and Michel Viso (CNES). BE, QC, MM, AT and IW are funded through the ERC Starter Grant ExoAI (GA 758892) and the STFC grants ST/P000282/1, ST/P002153/1, ST/S002634/1 and ST/T001836/1. NS acknowledges the support of the IRIS-OCAV, PSL. MP acknowledges support by the European Research Council under Grant Agreement ATMO 757858 and by the CNES. RB is a Ph.D. fellow of the Research Foundation--Flanders (FWO). WP and TZ have received funding from the European Research Council (ERC) under the European Union's Horizon 2020 research and innovation programme (grant agreement no. 679030/WHIPLASH). OV thanks the CNRS/INSU Programme National de Plan\'etologie (PNP) and CNES for funding support. GG acknowledges the financial support of the 2017 PhD fellowship programme of INAF.  LVM and DMG acknowledge the financial support of the ARIEL ASI grant n. 2018-22-HH.0.\\

\textbf{Software:} Iraclis \citep{Iraclis}, TauREx3 \citep{al-refaie_taurex3}, pylightcurve \citep{tsiaras_plc}, ExoTETHyS \citep{morello_exotethys}, ArielRad \citep{mugnai}, ExoWebb \citep{edwards_exowebb}, Astropy \citep{astropy}, h5py \citep{hdf5_collette}, emcee \citep{emcee}, Matplotlib \citep{Hunter_matplotlib}, Multinest \citep{multinest}, Pandas \citep{mckinney_pandas}, Numpy \citep{oliphant_numpy}, SciPy \citep{scipy}.

\appendix

TauREx3 employs Bayesian statistics as the cornerstone for the retrieval analysis \citep{Waldmann2015b, Waldmann2015a}. Bayes' theorem states that:

\begin{equation}
 P(\theta\mid x, \mathcal{M}) = \frac{P(x\mid\theta, \mathcal{M}) \, P(\theta, \mathcal{M})}{P(x\mid \mathcal{M})}
\end{equation}

\noindent where $P(\theta, \mathcal{M})$ is the Bayesian prior, $\mathcal{M}$ is the forward model. $P(\theta\mid x, \mathcal{M})$ is the posterior probability of the model parameters $ \theta$ given the data, \textit{x} assuming the forward model $\mathcal{M}$. Bayesian analysis is implemented in TauREx3 via nested sampling (NS).

TauREx3 includes the implementation of NS Bayesian statistics via Multinest \citep{Feroz_Hobson_2008, Feroz_2009, Feroz_2013}. NS employs Monte Carlo approach which constrains via ellipsoids encompassing the parameter space of the highest likelihood. NS determines the Bayesian evidence which is given by:

\begin{equation}
E = \int P(\theta \mid \mathcal{M})P(x\mid\theta, \mathcal{M})d\theta 
\end{equation}

\noindent where $P(x, \mathcal{M})$ is the evidence. These statistical products produced by Multinest are used to perform the best fit model selection. NS performed by Multinest also allows for efficient parallelisation which permits the usage of high performance cluster computing.

\begin{table*}
    \centering
    \begin{tabular}{ccccc}
    \hline \hline
    Wavelength [$\mu$m] & Bandwidth [$\mu$m] & Transit Depth [\%] & Eclipse Depth [\%] & Instrument\\\hline  \hline
    0.8 & 0.4 & 0.7961 $\pm$ 0.0018 & - & TESS\\
    1.12625 & 0.0219 & 0.8159 $\pm$  0.0045 & 0.03737 $\pm$ 0.0045 & HST WFC3 \\
    1.14775 & 0.0211 & 0.8134  $\pm$ 0.0045 & 0.03951 $\pm$ 0.0046 & HST WFC3 \\
    1.16860 & 0.0206 & 0.8062  $\pm$ 0.0048 & 0.03669 $\pm$ 0.0045 & HST WFC3 \\
    1.18880 & 0.0198 & 0.8152  $\pm$ 0.0048 & 0.03544 $\pm$ 0.0048 & HST WFC3 \\
    1.20835 & 0.0193 & 0.8195 $\pm$  0.0047 & 0.03876 $\pm$ 0.0041 & HST WFC3  \\
    1.22750 & 0.0190 & 0.8181 $\pm$  0.0043 & 0.03787 $\pm$ 0.0049 & HST WFC3  \\
    1.24645 & 0.0189 & 0.8114 $\pm$  0.0049 & 0.03865 $\pm$ 0.0043 & HST WFC3  \\
    1.26550 & 0.0192 & 0.8172 $\pm$  0.0054 & 0.03292 $\pm$ 0.0046 & HST WFC3  \\
    1.28475 & 0.0193 & 0.8170  $\pm$ 0.0060 & 0.04363 $\pm$ 0.0043 & HST WFC3 \\
    1.30380 & 0.0188 & 0.8130  $\pm$ 0.0055 & 0.04558 $\pm$ 0.0042 & HST WFC3 \\
    1.32260 & 0.0188 & 0.8237  $\pm$ 0.0046 & 0.03523 $\pm$ 0.0044 & HST WFC3 \\
    1.34145 & 0.0189 & 0.8163  $\pm$ 0.0050 & 0.03577 $\pm$ 0.0043 & HST WFC3 \\
    1.36050 & 0.0192 & 0.8174  $\pm$ 0.0042 & 0.04638 $\pm$ 0.0044 & HST WFC3 \\
    1.38005 & 0.0199 & 0.8117  $\pm$ 0.0043 & 0.04633 $\pm$ 0.0045 & HST WFC3 \\
    1.40000 & 0.0200 & 0.8171  $\pm$ 0.0047 & 0.04750 $\pm$ 0.0038 & HST WFC3 \\
    1.42015 & 0.0203 & 0.8191  $\pm$ 0.0047 & 0.04555 $\pm$ 0.0051 & HST WFC3 \\
    1.44060 & 0.0206 & 0.8127  $\pm$ 0.0056 & 0.04515 $\pm$ 0.0046 & HST WFC3 \\
    1.46150 & 0.0212 & 0.8179  $\pm$ 0.0050 & 0.04580 $\pm$ 0.0044 & HST WFC3 \\
    1.48310 & 0.0220 & 0.8164  $\pm$ 0.0047 & 0.04480 $\pm$ 0.0046 & HST WFC3 \\
    1.50530 & 0.0224 & 0.8006  $\pm$ 0.0048 & 0.05031 $\pm$ 0.0048 & HST WFC3 \\
    1.52800 & 0.0230 & 0.8112 $\pm$  0.0052 & 0.04825 $\pm$ 0.0048 & HST WFC3  \\
    1.55155 & 0.0241 & 0.8052 $\pm$  0.0052 & 0.04765 $\pm$ 0.0050 & HST WFC3  \\
    1.57625 & 0.0253 & 0.7989 $\pm$  0.0063 & 0.05397 $\pm$ 0.0052 & HST WFC3  \\
    1.60210 & 0.0264 & 0.7872 $\pm$  0.0057 & 0.04457 $\pm$ 0.0048 & HST WFC3  \\
    1.62945 & 0.0283 & 0.7936 $\pm$  0.0057 & 0.04332 $\pm$ 0.0051 & HST WFC3  \\
    3.6 & 0.75 & 0.7925$\pm$0.0062 & 0.1688 $\pm$ 0.0046 & Spitzer IRAC$^*$\\
    4.5 & 1.015 & 0.8092$\pm$0.0036 & 0.1896 $\pm$ 0.0057 & Spitzer IRAC$^*$\\ \hline 
    \multicolumn{5}{c}{$^*$Taken from  \cite{garhart_spitzer}.}\\\hline \hline
    \end{tabular}
\caption{Spectral data of KELT-7\,b used in this study.}
\end{table*}

\begin{figure*}[p]
    \centering
    \includegraphics[width = 0.95\textwidth]{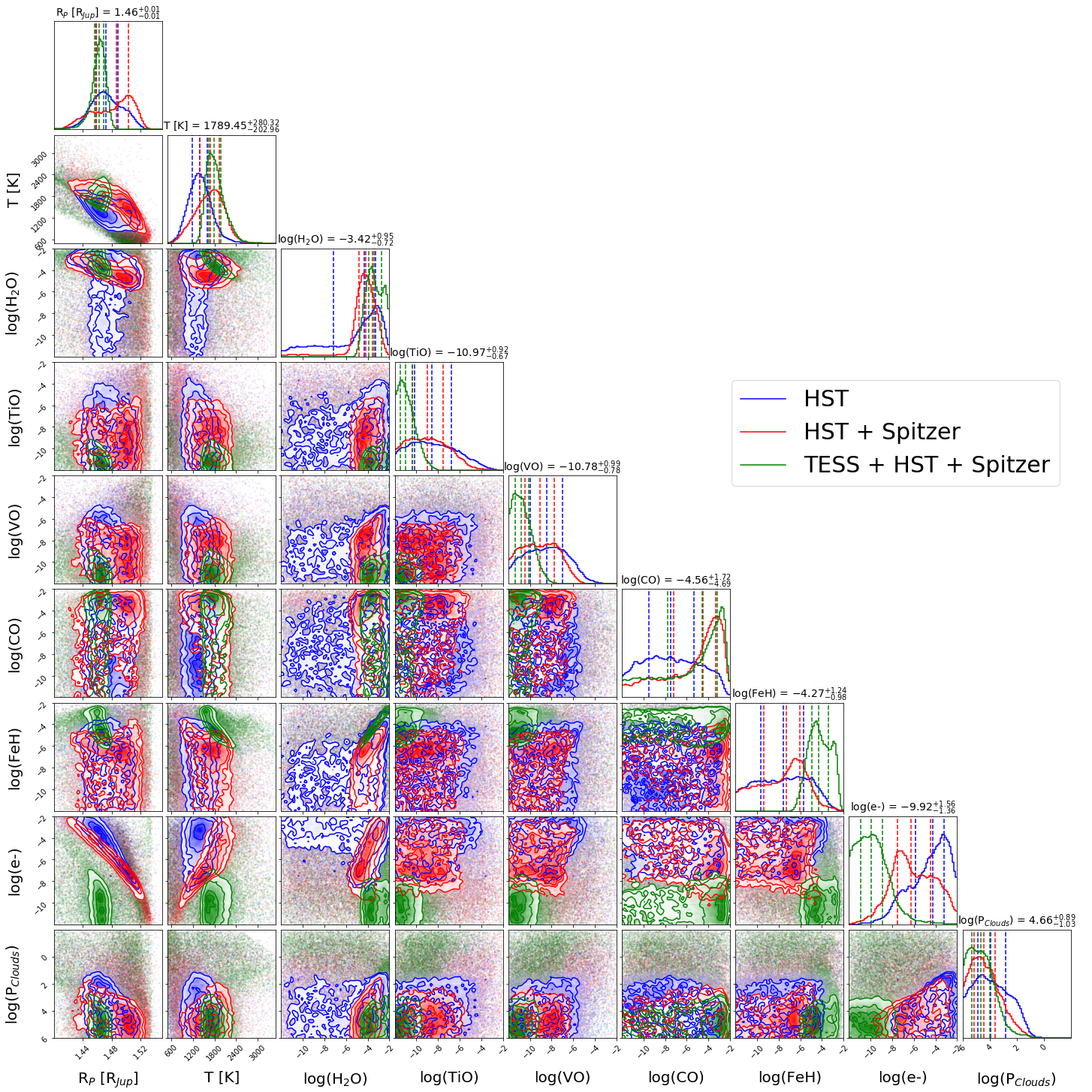}
    \caption{Posterior distributions for atmospheric retrievals of KELT-7\,b with various datasets. The addition of Spitzer data brings little change to the atmospheric properties while TESS data drives the retrieval to favour FeH over H-. We note that this FeH abundance is far above the expected (log(FeH)$\sim -7$). The reported values for each parameter are those obtained with the fit on all three datasets.}
    \label{fig:posteriors_trans}
\end{figure*}

\begin{figure*}[p]
    \centering
    \includegraphics[width = 0.95\textwidth]{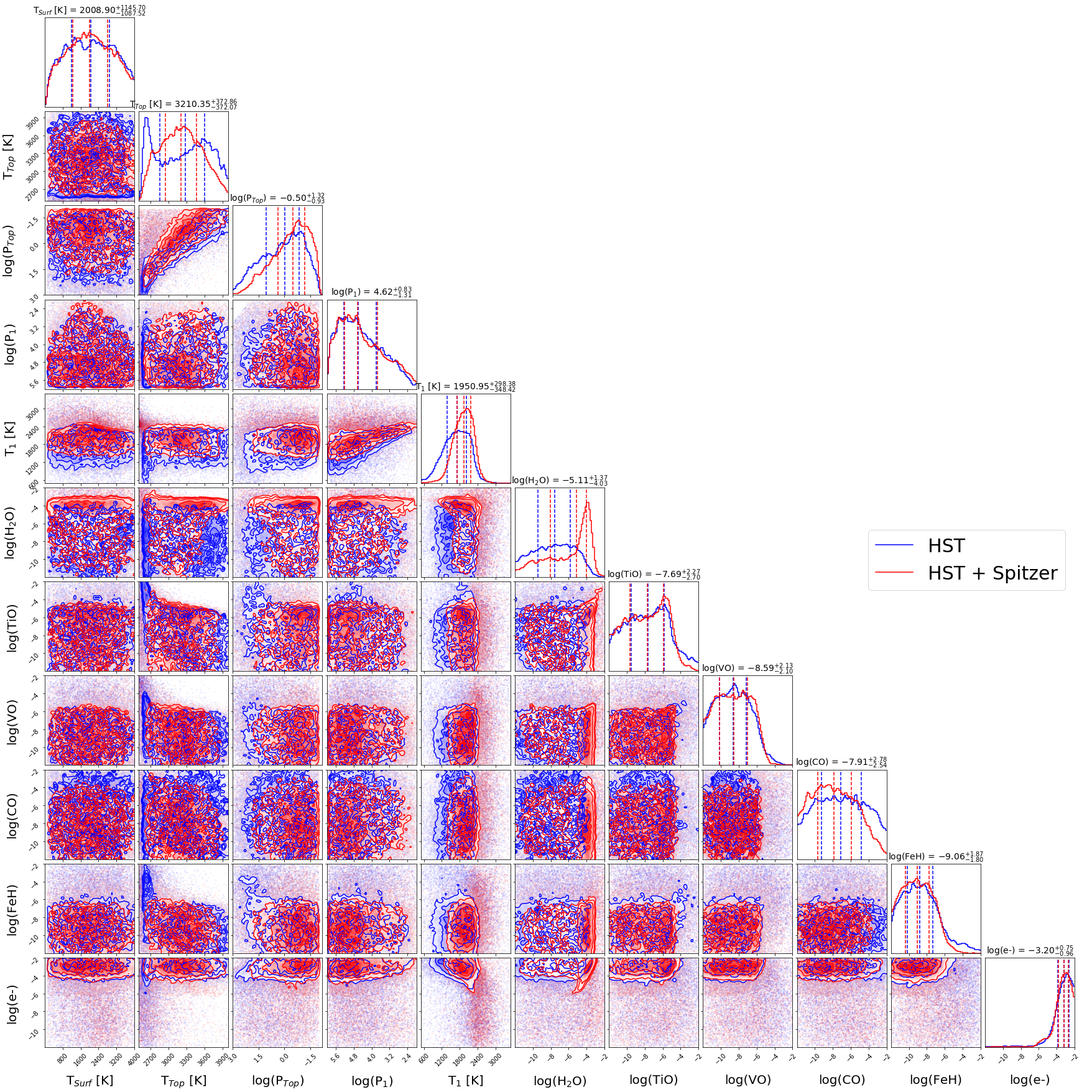}
    \caption{Posterior distributions for atmospheric retrievals of the emission spectra of KELT-7\,b with and without Spitzer data (blue and red respectively). The addition of Spitzer data brings little change to the atmospheric properties except the derived water abundance which appears clearly with Spitzer data. The reported values for each parameter are those obtained with the fit on both datasets.}
    \label{fig:posteriors_ec}
\end{figure*}

\begin{figure*}[p]
    \centering
    \includegraphics[width = 0.95\textwidth]{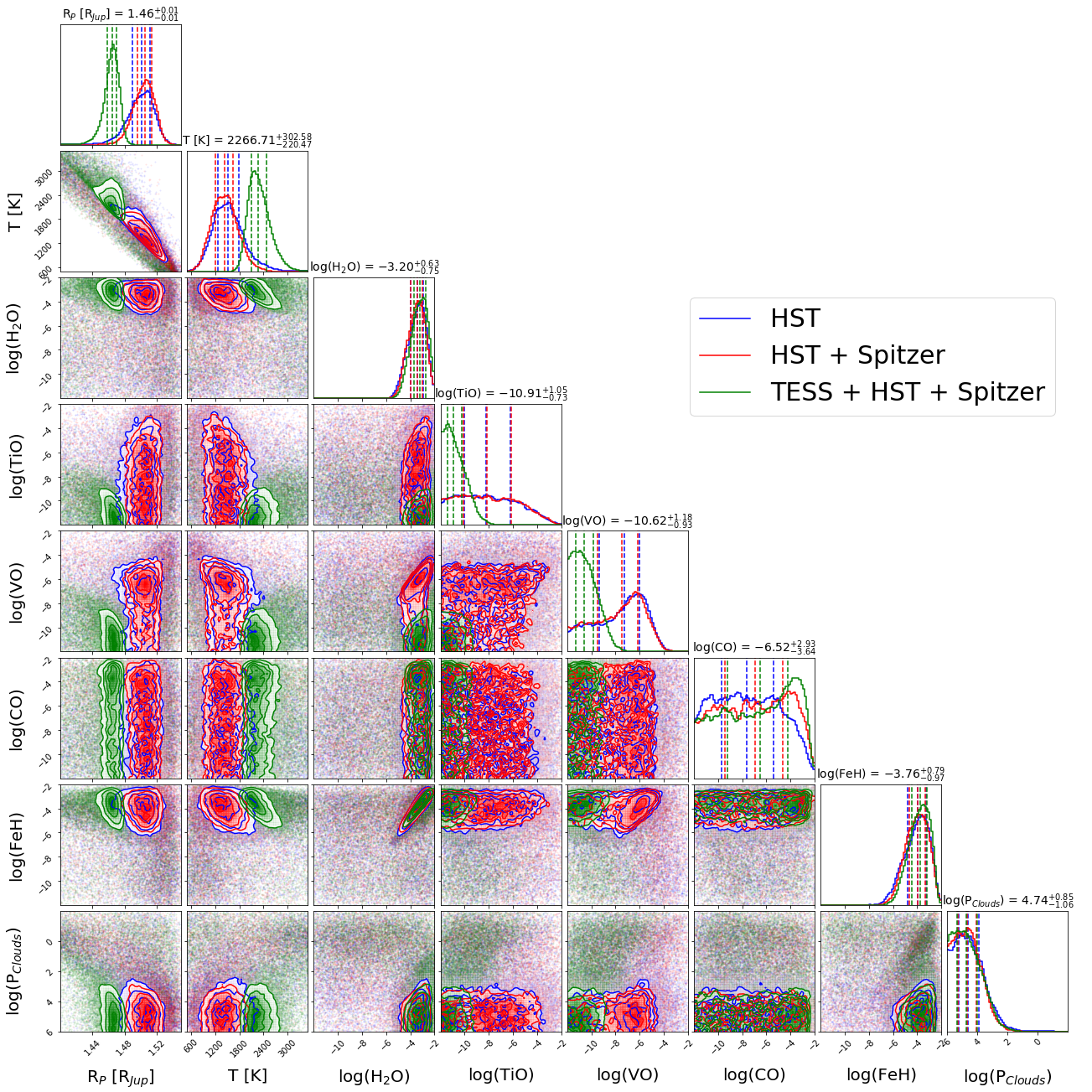}
    \caption{Posterior distributions for atmospheric retrievals of the transmission spectra of KELT-7\,b with various datasets, this time without including H- as an opacity source. The addition of Spitzer data again has little effect on the retrieved atmospheric properties while TESS data drives the retrieval to higher temperatures and rules out the presence of TiO and VO. All cases require an FeH abundance which is far greater than expected. The reported values for each parameter are those obtained with the fit on all three datasets.}
    \label{fig:posteriors_nohminus}
\end{figure*}

\clearpage

\begin{figure}
    \centering
    \includegraphics[width=0.45\textwidth]{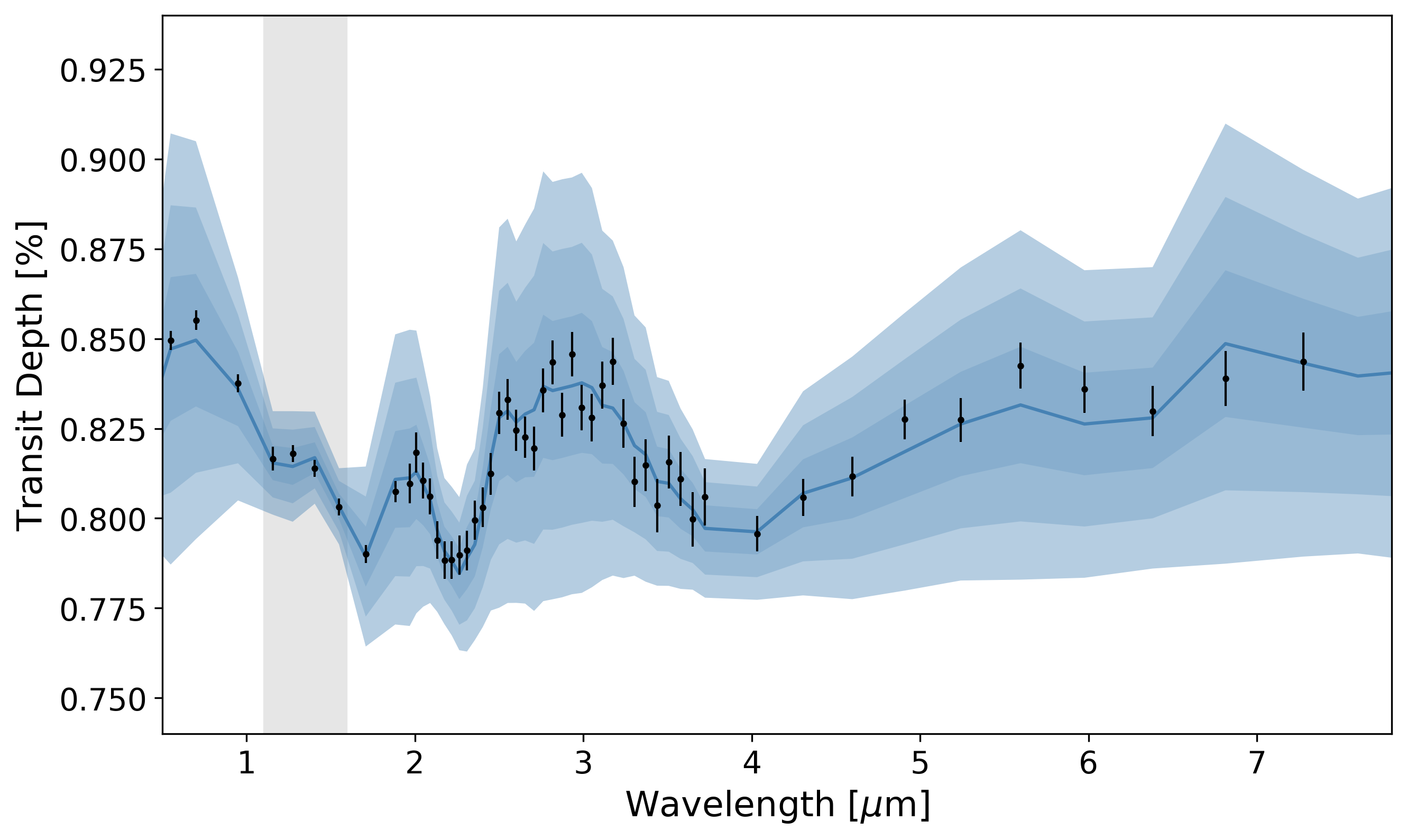}
    \includegraphics[width=0.45\textwidth]{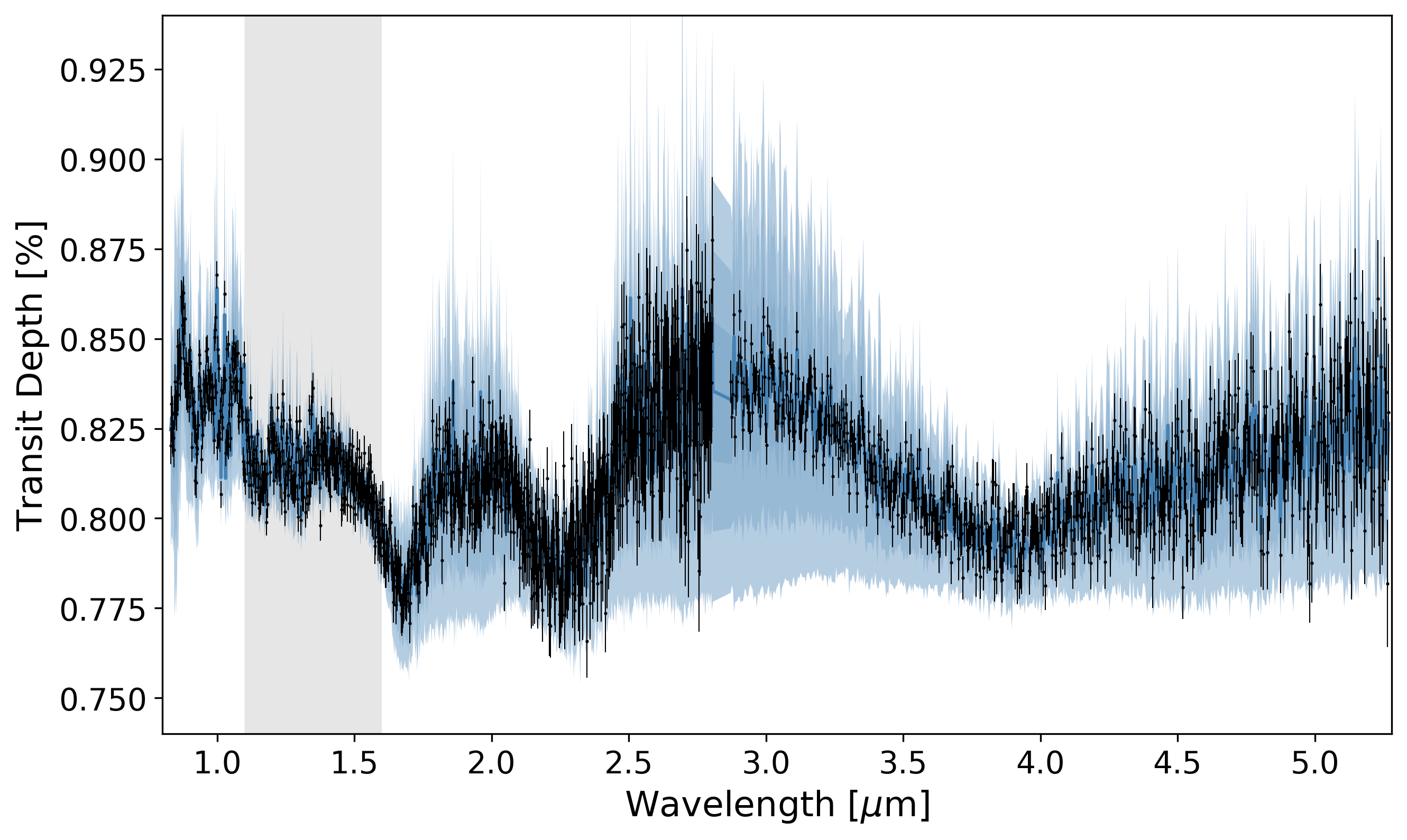}
    \includegraphics[width=0.45\textwidth]{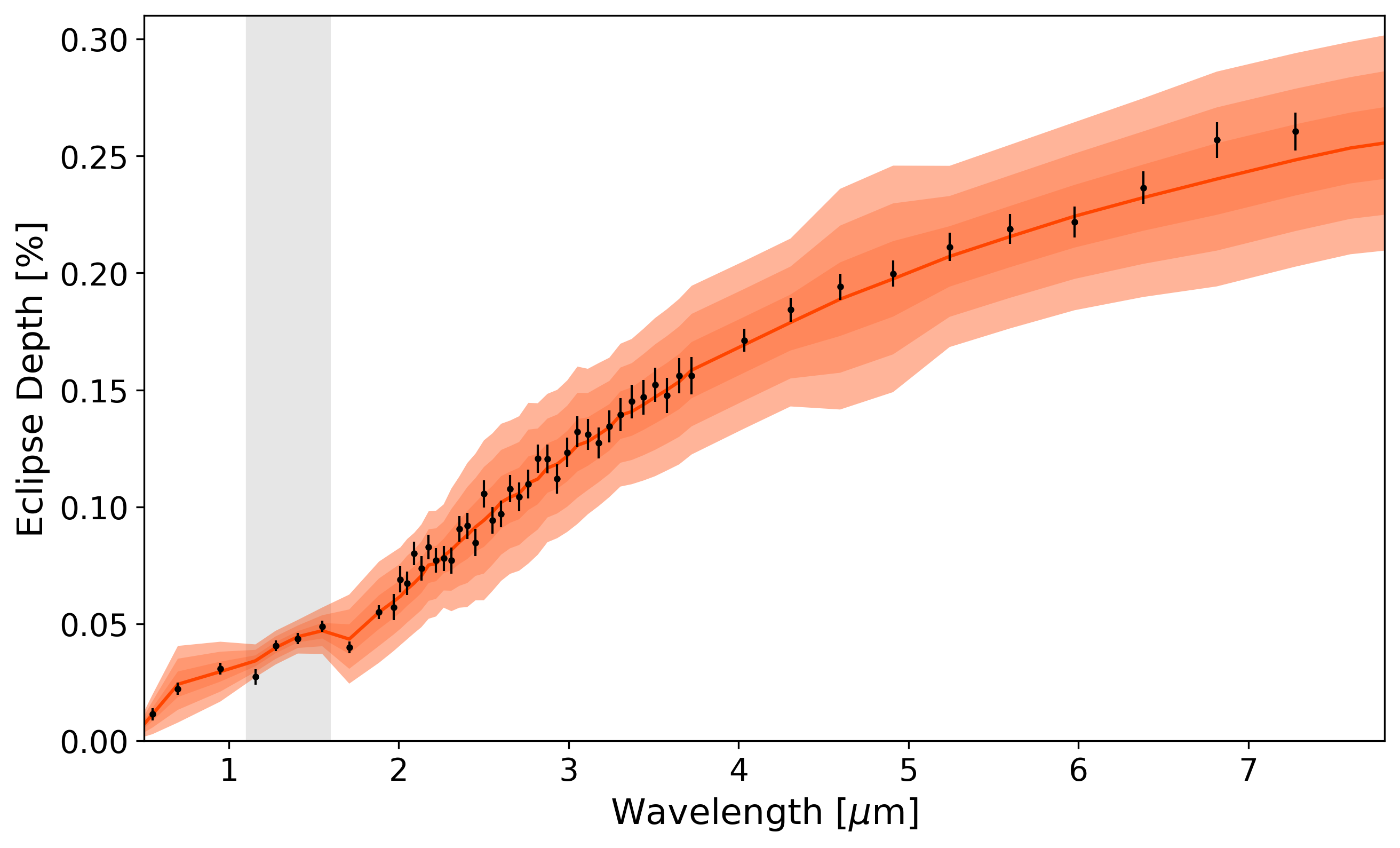}
    \includegraphics[width=0.45\textwidth]{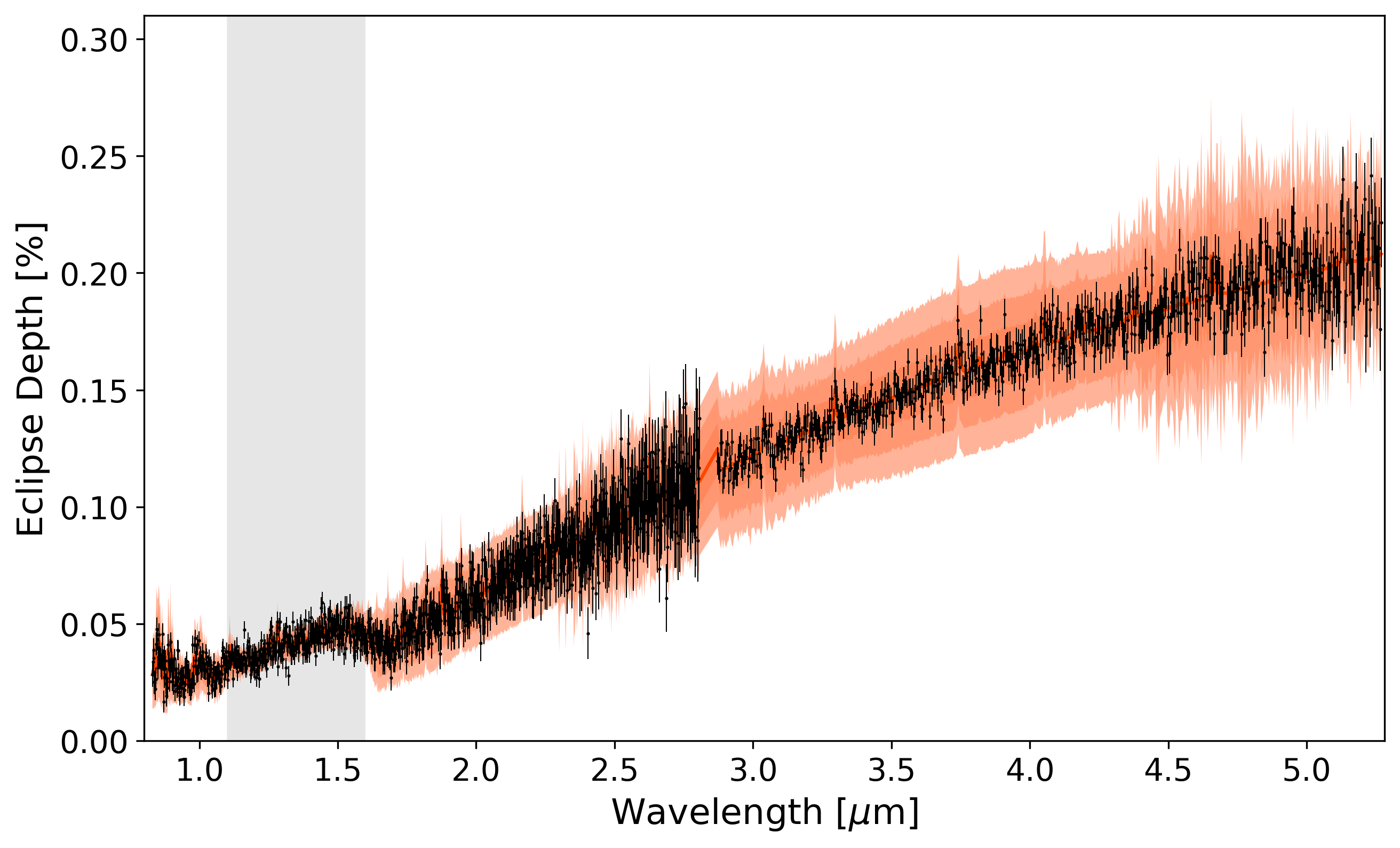}
    \caption{Simulated Ariel and JWST observations of the best-fit solutions retrieved in this work. The top panels show transmission spectra while the bottom panels show emission spectra. For Ariel, 2 stacked observations have been assumed while for JWST we modelled a single observation with NIRISS GR700XD as well as an observation with NIRSpec G395M.}
    \label{fig:simulations}
\end{figure}


\bibliographystyle{yahapj}
\bibliography{ms1}

\end{document}